\documentclass[11pt,a4paper]{article}
\pdfoutput=1
\usepackage{jheppub}
\usepackage{cite}
\usepackage{graphicx}
\usepackage{longtable}
\usepackage{wrapfig}
\usepackage{rotating}
\usepackage[normalem]{ulem}
\usepackage{amsmath}
\usepackage{amssymb}
\usepackage{capt-of}
\usepackage{hyperref}
\usepackage{mathrsfs}
\usepackage{bbold}
\usepackage{cancel}
\usepackage{braket}
\usepackage{tikz}
\usepackage{lscape}
\usepackage{setspace}
\usepackage{amsmath,amsfonts,epsfig}
\usepackage{tcolorbox}
\usepackage{cite}
\usepackage{subcaption}
\usepackage{xstring}
\usepackage{braket}

\usepackage{physics}
\usepackage[normalem]{ulem}
\usepackage{braket}
\usepackage{caption}
\usepackage{cancel}
\usepackage{subcaption}
\usepackage{multirow, graphicx,amssymb,url,mathrsfs,amsmath}
\usepackage{wrapfig,boxedminipage,epsfig}
\usepackage{amsxtra,amstext,latexsym,dsfont,amsfonts}
\usepackage{color}
\usepackage{tcolorbox}
\usepackage{float}
\usepackage{slashed}
\usepackage{calligra}
\usepackage{enumerate}
\usepackage{mathtools}
\DeclareFontShape{T1}{calligra}{m}{n}{<->s*[2.2]callig15}{}
\DeclareMathAlphabet{\mathcalligra}{T1}{calligra}{m}{n}

\graphicspath{{./figs/}}

\newcommand\GM[1][]{{\rm GM}^{\IfInteger{#1}{(#1)}{\mathtt{(#1)}}}}
\renewcommand\S[1][]{S^{\IfInteger{#1}{(#1)}{\mathtt{(#1)}}}}
\newcommand\A[1][]{\mathcal{A}^{\IfInteger{#1}{(#1)}{\mathtt{(#1)}}}}

\newcommand{\be}{\begin{equation}}
\newcommand{\ee}{\end{equation}}

\title{Structural Obstruction to Replica Symmetry Breaking for Multi-Entropy in Random Tensor Networks}

\author[a]{Sriram Akella} 

\author[b, c]{Norihiro Iizuka}

\affiliation[a]{\it Department of Theoretical Physics, Tata Institute of Fundamental Research, Homi Bhabha Road,
Colaba, Mumbai 400005}
 
\affiliation[b]{\it Department of Physics, National Tsing Hua University, Hsinchu 300044, Taiwan}

\affiliation[c]{\it Yukawa Institute for Theoretical Physics, Kyoto University, Kyoto 606-8502, Japan}

\emailAdd{sriram.akella@tifr.res.in}
\emailAdd{iizuka@phys.nthu.edu.tw}

\abstract{
We study replica symmetry breaking (RSB) for multi-entropy in the random-tensor-network (RTN) domain-wall spin model. Our main result is that, within this framework, multi-entropy has a structural obstruction to RSB for any R\'enyi index~$n$ and any multipartite number~$\mathtt{q}$. This obstruction arises because the boundary permutations relevant to multi-entropy are organized along mutually incompatible coordinate directions of the replica hypercube, and therefore do not admit a nontrivial common geodesic intermediate permutation~$\tau$ in the Cayley graph of~$S_N$. This is in sharp contrast to entanglement negativity, which does admit such a $\tau$-mediated saddle and exhibits RSB in the same framework. As a robustness check, we also consider a toy $\mathbb{Z}_2$ gauge extension of the spin model with a minimal bulk gauge constraint. Numerical evidence in this gauged model indicates that multi-entropy continues to show no sign of RSB at $n=2$ and $n=3$, while negativity continues to exhibit RSB. Our results show that, within the RTN spin-model description, multi-entropy is not ``RSB-friendly'': its boundary data are structurally incompatible with a nontrivial common geodesic intermediate permutation, unlike negativity.}

\begin{document}
\maketitle

\section{Introduction}

Quantum entanglement plays a central role in holography.
So far, much of the progress in understanding its bulk interpretation has focused on \emph{bipartite} entanglement entropy.
In that setting, the Ryu--Takayanagi formula~\cite{Ryu:2006bv,Ryu:2006ef} and its extensions~\cite{Hubeny:2007xt} provide a remarkably clear geometric picture, relating boundary entanglement entropy to extremal surfaces in the bulk.

However, quantum states generically contain intrinsically multipartite entanglement as well.
It is therefore natural to ask whether multipartite entanglement admits an equally meaningful holographic interpretation, and if so, what the corresponding bulk picture should be.
This question is especially important if one wishes to move beyond bipartite diagnostics and understand the structure of entanglement in holography in a more complete way.

A useful toy framework for probing such questions is provided by the random-tensor-network (RTN) domain-wall spin model introduced by Hayden et al.~\cite{Hayden:2016cfa}.
In this framework, replicated partition functions are mapped to an effective domain-wall problem whose spin variables take values in a permutation group, and in the bipartite case the dominant contribution is governed by a minimal-cut structure analogous to the Ryu--Takayanagi saddle. This makes the RTN domain-wall spin model a useful toy model for exploring possible bulk geometric descriptions of entanglement observables. 
More generally, the same framework has also proved useful for analyzing replicated observables beyond bipartite entanglement entropy, including cases such as negativity where nontrivial replica saddles can arise~\cite{Dong:2021clv}, and reflected entropy \cite{Akers:2021pvd, Akers:2022zxr, Akers:2024pgq}.

A key issue in this context is whether the dominant relevant replica saddles preserve replica symmetry or are replaced by nontrivial intermediate configurations.
In other words, one would like to understand when the dominant replica saddle is the naive symmetric one, namely the one obtained by directly extending the boundary-imposed permutation sectors into the bulk, and when it is instead replaced by a nontrivial $\tau$-mediated saddle.
This is precisely the question of replica symmetry breaking (RSB), and it provides an important diagnostic for the structure of possible bulk interpretations of multipartite observables.

A particularly interesting question is whether the R\'enyi multi-entropy $S_n^{(\mathtt{q})}$, a recently proposed multipartite entanglement measure,\footnote{While multi-entropy is a symmetric $\mathtt{q}$-partite entanglement measure, it does not by itself isolate genuinely $\mathtt{q}$-partite correlations. To extract the genuine $\mathtt{q}$-partite contribution, one should instead consider genuine multi-entropy, obtained by subtracting lower-partite contributions from multi-entropy~\cite{Iizuka:2025ioc,Iizuka:2025caq}. The mathematical structure of this construction was recently clarified by Gadde~\cite{Gadde:2026msg} in terms of M\"obius inversion on the partition lattice.}
exhibits RSB in this framework~\cite{Gadde:2022cqi,Penington:2022dhr,Gadde:2023zzj}. This question is especially important because multi-entropy has been proposed to admit a simple bulk geometric interpretation in terms of a soap-film-like configuration~\cite{Gadde:2022cqi}.
At the same time, its definition relies on a replica construction and an eventual continuation away from integer R\'enyi index, whose validity is not automatic. For example, in gravitational replica calculations, the tripartite R\'enyi multi-entropy spontaneously breaks bulk replica symmetry for R\'enyi index $n \geq 3$ \cite{Penington:2022dhr}, obstructing a naive application of the Lewkowycz-Maldacena argument towards $n = 1$ \cite{Lewkowycz:2013nqa}.

In the example of entanglement negativity, the dominant saddle in the RTN domain-wall spin model breaks replica symmetry \cite{Dong:2021clv}, consistent with expectations for general holographic states \cite{Dong:2024gud}. Does the multi-entropy too conform to our expectations from holography by breaking replica symmetry for R\'enyi index $n \geq 3$ in the RTN domain-wall spin model?

Our main result is that, in the RTN domain-wall spin model,  multi-entropy has a structural obstruction to RSB, for any R\'enyi index $n$ and any multipartite number $\mathtt{q}$.
Rather than being an accident of a few examples, this turns out to reflect a structural feature of the boundary permutations relevant to multi-entropy.
The replica labels are naturally organized as points on a replica hypercube, while the boundary permutations act cyclically along different coordinate directions.
As a result, these boundary data do not admit a nontrivial common geodesic intermediate permutation in the Cayley graph of $S_N$.
In this sense, multi-entropy is structurally not ``RSB-friendly''.
This is in sharp contrast to the negativity example~\cite{Dong:2021clv}, which does admit such a nontrivial intermediate permutation and exhibits RSB in the same framework.
Thus, even within a common effective spin-model description, negativity and multi-entropy behave qualitatively differently.

This contrast is itself important, but it also raises the question of how far the RTN domain-wall spin model can be expected to capture the full replica structure of gravitational entanglement.
For example, it is most directly related to \emph{fixed-area states} of gravity, and in this sense it naturally captures the leading-order entanglement structure of gravity~\cite{Dong:2018seb}.
However, this does not by itself guarantee that it captures finer effects beyond the fixed-area regime at leading order in $1/G_N$.
Moreover, when interpreted as a holographic error-correcting code, it has trivial area operators~\cite{Harlow:2016vwg}.
More generally, if one wishes to move closer to gravitational systems, it is natural to consider ingredients such as gauge redundancy and global constraints, which are closely tied to diffeomorphism invariance.
In such settings, Hilbert-space factorization can be subtle, and additional structures such as Gauss-law constraints and boundary superselection sectors may become relevant.
It is therefore natural to ask whether the structural obstruction to RSB found above remains robust once one moves beyond the simplest RTN domain-wall spin model setting and incorporates such gauge-theoretic structure.

Motivated by this, we next consider a simple gauge extension of the RTN domain-wall spin model, following the general perspective of~\cite{Akers:2024ixq} (see also~\cite{Akers:2024wab,Dong:2023kyr,Qi:2022lbd, Donnelly:2016qqt}).  
Although the construction can be formulated for any finite group\footnote{Our construction can be generalized to certain non-compact gauge groups using the ideas developed in~\cite{Balasubramanian:2025rcr}.}, the resulting model is much less tractable analytically than the original RTN domain-wall spin model.
We therefore focus on the simple case $G=\mathbb{Z}_2$, which allows for explicit numerical analysis using the techniques of~\cite{Akella:2026xza}.
Our goal is not to claim that this simple model captures the full gravitational problem.
Rather, we use it as a first step toward incorporating gauge redundancy and global constraints into the RTN domain-wall spin-model framework.
In this toy gauged model, we still find no evidence of RSB for tripartite multi-entropy at $n=2$ and $n=3$, while negativity continues to exhibit RSB in the same setup.
This suggests that multi-entropy is significantly less prone to replica symmetry breaking than negativity, and that the absence of RSB is more robust than one might have expected from the simplest RTN domain-wall spin-model framework alone.

The rest of this paper is organized as follows.
In Sec.~\ref{sec:review} we review the RTN domain-wall spin model and recall the negativity example as a reference case where RSB does occur.
In Sec.~\ref{sec:noRSB} we present the main result: the structural obstruction to RSB for multi-entropy, first for $n=2$ as a warm-up, and then for general $n$ and $q$ from the structural properties of the boundary permutations.
In Sec.~\ref{sec:gauge} we introduce the toy $\mathbb{Z}_2$ gauge extension and present numerical evidence for the persistence of no-RSB.
In Sec.~\ref{sec:discussion} we summarize our results and discuss their implications.

\section{Review of the RTN domain-wall spin model and the negativity example}
\label{sec:review}

In this section, we review the RTN domain-wall spin model introduced by Hayden et al.~\cite{Hayden:2016cfa} and recall the negativity example, which provides a basic reference case where replica symmetry breaking (RSB) occurs \cite{Dong:2021clv}.
Our purpose here is to fix notation, define the two competing saddle types, and state the energetic criterion that will later be applied to multi-entropy. We will also emphasize the geometric difference between the naive $Y$-saddle and a nontrivial $\tau$-mediated saddle.

\subsection{Replica spin variables and the domain-wall Hamiltonian}

Random tensor networks~\cite{Hayden:2016cfa} prepare boundary states for which the replica trick can be applied to calculate quantum information-theoretic quantities like the entanglement entropy. The calculation is mapped to an effective classical spin model whose spin variable at each site takes values in the permutation group
\begin{equation}
  g_x \in S_N,
\end{equation}
where $N$ is the replica number. Which information-theoretic quantity is being calculated dictates the boundary conditions of the spin model, and in the large bond dimension limit, the dominant saddle is given by the minimum energy spin configuration in the bulk. The technical details of this mapping are reviewed in Appendix \ref{sec:app-crtn}.

 The interaction between neighboring sites depends only on the relative permutation
$g_x^{-1}g_y$.
More precisely, the bond Hamiltonian is
\begin{equation}
  H_{\langle xy\rangle}(g_x,g_y)
  =
  J\, d(g_x,g_y),
  \qquad
  d(g_x,g_y)\equiv N-C(g_x^{-1}g_y),
  \label{eq:bondhamiltonianreview}
\end{equation}
where $C(\sigma)$ denotes the number of cycles of $\sigma\in S_N$, including fixed points, and $J>0$ is a ferromagnetic coupling.

The quantity $d(g_x,g_y)$ is the Cayley distance with respect to transposition generators; equivalently, it is the minimal number of transpositions needed to relate $g_x$ and $g_y$. In particular,
\begin{equation}
  d(g_x,g_y)=0 \qquad \Leftrightarrow \qquad g_x=g_y,
\end{equation}
so the model energetically favors aligned permutations. The replicated problem is therefore reduced to a domain-wall problem in which different permutation domains compete through interface tensions determined by $d(g_x,g_y)$.

It is useful to keep a few simple examples in mind.
For $N=2$, one has
\begin{equation}
  S_2=\{e,(12)\},
\end{equation}
so the Cayley distance takes only the values $0$ and $1$.
The model is then essentially a two-state ferromagnetic nearest-neighbor Ising model.
For example, in the bipartite second R\'enyi entanglement entropy, namely the purity $\Tr \rho_A^2$, the boundary condition is $(12)$ on $A$ and $e$ on $\bar A$. This follows from the replica construction of $\Tr\rho_A^2$: on $\bar A$, each replica is contracted with itself, giving the identity permutation $e$, whereas on $A$, adjacent replicas are cyclically glued, giving the transposition $(12)$. More generally, for $\Tr\rho_A^N$, one obtains the cyclic permutation $(1\,2\,\cdots\,N)$ on $A$ and $e$ on $\bar A$. The corresponding saddle-point problem is therefore a domain-wall problem separating an $e$ region from a $(12)$ region, and the dominant contribution is given by the minimal such interface.

For $N=3$, one has
\begin{equation}
  S_3=\{e,(12),(13),(23),(123),(132)\},
\end{equation}
and the Cayley distance takes the values
\begin{equation}
  d=0,1,2,
\end{equation}
according to whether the relative permutation is the identity, a transposition, or a $3$-cycle.
For example, the bipartite third R\'enyi entropy $\Tr \rho_A^3$ is again reduced to a domain-wall problem with boundary conditions $(123)$ on $A$ and $e$ on $\bar A$. Unlike the $N=2$ case, however, the group now contains additional nontrivial elements such as $(12)$, which can in principle lie between distinct boundary sectors in the Cayley geometry\footnote{For the bipartite third R\'enyi entropy, however, this does not by itself imply a distinct saddle with an intermediate permutation. The dominant configuration is still the minimal-length domain wall separating the $e$ and $(123)$ boundary sectors, while an element such as $(12)$ can at most appear along the same interface between $e$ and $(123)$.}.

More generally, the advantage of the domain-wall formulation is that the competition among replica saddles is reduced to a geometric problem. If two neighboring domains carry permutations $g$ and $h$, then the corresponding interface tension is proportional to $d(g,h)$. The dominant saddle is therefore obtained by minimizing the total domain-wall energy subject to the boundary conditions imposed by the observable under study. A central question then is whether a nontrivial intermediate permutation can form an independent competing domain in the bulk. If such a saddle dominates over the naive one, we regard this as replica symmetry breaking because the dominant bulk configuration is no longer the one obtained by directly extending the boundary-imposed permutation sectors into the bulk, but instead involves an additional permutation domain that is not present in the boundary data.

\subsection{Boundary conditions for tripartite measures}

So far, we have reviewed the domain-wall spin model and its basic application to bipartite entanglement measures. We now turn to multi-partite quantum measures, focusing first on the tripartite case. For this purpose, we consider a disk geometry whose boundary is divided into three arcs, denoted by $A$, $B$, and $C$. On these arcs we impose
\begin{equation}
  g(\theta)=
  \begin{cases}
    g_A, & \theta \in A,\\
    g_B, & \theta \in B,\\
    g_C, & \theta \in C,
  \end{cases}
\end{equation}
for some prescribed permutations $g_A,g_B,g_C\in S_N$. The choice of these boundary permutations depends on the multi-partite quantum measure under consideration.

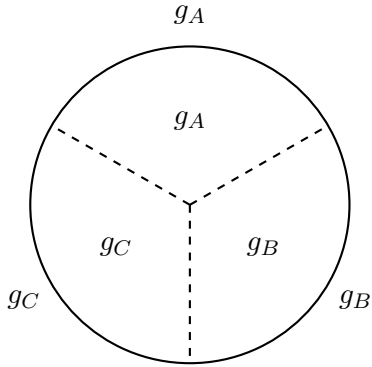
\begin{figure}[t]
  \centering
  \makebox[\linewidth][c]{%
    \begin{tikzpicture}[scale=1.4]
      \draw[thick] (0,0) circle (1.5);

      \coordinate (O) at (0,0);

      \draw[thick, dashed] (O) -- (150:1.5);
      \draw[thick, dashed] (O) -- (30:1.5);
      \draw[thick, dashed] (O) -- (270:1.5);

      \node at (90:1.8)   {$g_A$};
      \node at (330:1.8)  {$g_B$};
      \node at (210:1.8)  {$g_C$};

      \node at (90:0.8)   {$g_A$};
      \node at (330:0.8)  {$g_B$};
      \node at (210:0.8)  {$g_C$};
    \end{tikzpicture}%
  }
  \caption{Naive $Y$-configuration for tripartite boundary conditions.}
  \label{fig:naive-Ygeneric}
\end{figure}

A natural candidate saddle is the naive $Y$-configuration, in which the three
boundary-imposed phases meet directly in the bulk at a single trivalent junction as shown in Fig.~\ref{fig:naive-Ygeneric}. 

Its energy is controlled by the three pairwise tensions
\begin{equation}
  d(g_A,g_B),\qquad d(g_B,g_C),\qquad d(g_C,g_A).
\end{equation}
We therefore define
\begin{equation}
  S_{\rm pair}
  :=
  d(g_A,g_B)+d(g_B,g_C)+d(g_C,g_A).
  \label{eq:Spairreview}
\end{equation}

If one approximates the bulk geometry by flat space, and takes the three arms of the $Y$ to have comparable length $L$, then the energy of this naive saddle is schematically
\begin{equation}
  E_{(a)} \sim J\,L\, S_{\rm pair}.
\end{equation}

This flat-space estimate is, however, only heuristic. In the actual holographic setup, the bulk geometry is AdS rather than flat space, so the domain-wall lengths are not characterized by a single finite scale $L$.

In particular, in AdS$_3$ each arm stretches toward the asymptotic boundary, and its proper length acquires the usual UV divergence,
\begin{equation}
  \ell \sim \log \frac{1}{\epsilon},
  \qquad (\epsilon \to 0),
\end{equation}
with $\epsilon$ the boundary cutoff. Accordingly, the energy is more properly controlled by the sum of geodesic lengths weighted by the corresponding pairwise tensions, rather than by the flat-space expression $JLS_{\rm pair}$. Schematically, one then expects
\begin{equation}
  E_{(a)}
  \sim
  J\,\log\frac{1}{\epsilon}\, S_{\rm pair}
  + E_{(a)}^{\rm finite},
\end{equation}
where $E_{(a)}^{\rm finite}$ denotes the cutoff-independent finite part.

However, the bulk spins are allowed to explore the full group $S_N$, so an
additional intermediate permutation $\tau\in S_N$ may enter.
This leads to a competing $\tau$-mediated configuration in which a central bulk
region is occupied by a single permutation $\tau$, separated from the boundary
phases $g_A,g_B,g_C$ by three interfaces.
The relevant quantity is then
\begin{equation}
  S_\tau
  :=
  d(g_A,\tau)+d(g_B,\tau)+d(g_C,\tau).
  \label{eq:Staureview}
\end{equation}

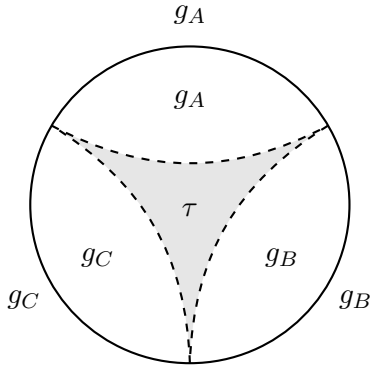
\begin{figure}[t]
  \centering
  \makebox[\linewidth][c]{%
    \begin{tikzpicture}[scale=1.4]
      \draw[thick] (0,0) circle (1.5);

      \coordinate (P1) at (150:1.5);
      \coordinate (P2) at (30:1.5);
      \coordinate (P3) at (270:1.5);

      \fill[gray!20]
        (P1) to[bend right=28] (P2)
             to[bend right=28] (P3)
             to[bend right=28] (P1);

      \draw[thick, dashed]
        (P1) to[bend right=28] (P2);
      \draw[thick, dashed]
        (P2) to[bend right=28] (P3);
      \draw[thick, dashed]
        (P3) to[bend right=28] (P1);

      \node at (90:1.8)   {$g_A$};
      \node at (330:1.8)  {$g_B$};
      \node at (210:1.8)  {$g_C$};

      \node at (90:1.0)   {$g_A$};
      \node at (330:1.0)  {$g_B$};
      \node at (210:1.0)  {$g_C$};

      \node at (0,-0.05) {$\tau$};
    \end{tikzpicture}%
  }
  \caption{$\tau$-mediated configuration with a central $\tau$ domain bounded by RT-like curved interfaces.}
  \label{fig:tau-mediated-curved}
\end{figure}

If one approximates the bulk geometry by flat space, and models the $\tau$-mediated configuration by an inscribed equilateral triangle, then its energy is schematically
\begin{equation}
  E_{(b)} \sim \sqrt{3}\,J\,L\,S_\tau,
\end{equation}
since the perimeter of an inscribed equilateral triangle is $\sqrt{3}$ times the disk radius.

As before, however, this flat-space estimate is only heuristic. In the actual holographic setup, the three interfaces are geodesic arcs in AdS and acquire the same near-boundary UV-divergent contribution discussed above. Thus the energy is more properly determined by the total weighted length of the three AdS geodesic segments bounding the $\tau$ region, rather than by the flat-space expression $\sqrt{3}JLS_\tau$. One therefore expects, schematically,
\begin{equation}
  E_{(b)}
  \sim
  J\,\log\frac{1}{\epsilon}\,S_\tau
  + E_{(b)}^{\rm finite},
\end{equation}
where $E_{(b)}^{\rm finite}$ denotes the cutoff-independent finite part.

Therefore, the leading competition between the two saddles is governed first by the comparison between $S_{\rm pair}$ and $S_\tau$. Only when these leading divergent contributions are tied does one need to compare the finite parts $E_{(a)}^{\rm finite}$ and $E_{(b)}^{\rm finite}$. The competition between the naive $Y$-saddle and the $\tau$-mediated saddle is the basic mechanism behind replica symmetry breaking in this framework. In the negativity example, this competition is realized explicitly, and the $\tau$-mediated saddle can dominate over the naive $Y$-configuration.

\subsection{Operational criterion for replica symmetry breaking}

In this paper we use the following operational definition.
Suppose that
\begin{equation}
  \tau \notin \{g_A,g_B,g_C\},
\end{equation}
and that a $\tau$-mediated configuration is energetically preferred over the
naive $Y$-saddle.
Then we say that replica symmetry is broken.

In the flat-space approximation discussed above, the condition for a genuinely
$\tau$-mediated saddle to dominate is
\begin{equation}
  S_{\rm pair} \ge \sqrt{3}\, S_\tau .
  \label{eq:flatcriterionreview}
\end{equation}
This is the natural condition obtained by comparing a trivalent $Y$ network with a $\tau$-mediated configuration modeled by an inscribed equilateral triangle, whose perimeter is $\sqrt{3}$ times the disk radius.

In AdS, the situation is slightly different.
The leading contribution to the domain-wall energy is dominated by the near-boundary region.
A naive $Y$-configuration has three UV legs, while each edge of a triangular
$\tau$ domain is anchored at two boundary endpoints.
As a result, the leading comparison becomes
\begin{equation}
  S_{\rm pair} \ge 2\, S_\tau .
  \label{eq:AdScriterionreview}
\end{equation}
This is the AdS criterion for the $\tau$-mediated saddle to compete with the
naive $Y$-saddle.

A useful general fact is that the Cayley distance is a genuine metric and therefore obeys the triangle inequality. Applying it to the three pairs $(g_A,g_B)$, $(g_B,g_C)$, and $(g_C,g_A)$ with intermediate point $\tau$, one finds
\begin{equation}
  d(g_A,g_B)\le d(g_A,\tau)+d(\tau,g_B),
\end{equation}
and similarly for the other two pairs. Summing these inequalities gives
\begin{equation}
  S_{\rm pair}\le 2S_\tau .
  \label{eq:universalboundreview}
\end{equation}
Therefore, in AdS, the condition \eqref{eq:AdScriterionreview} can only be satisfied when
\begin{equation}
  S_{\rm pair}=2S_\tau .
  \label{eq:saturationreview}
\end{equation}
That is, each of the triangle inequalities must be saturated separately, so $\tau$ must lie on a geodesic between each pair of boundary permutations. In this sense, $\tau$ plays the role of a common intermediate point in the Cayley geometry. This criterion will be useful in the next section.

\subsection{Negativity as a review example}

Let us now recall the standard negativity-type boundary condition in the
domain-wall analysis, following \cite{Dong:2021clv}.
We begin with the $N=3$ case, which already captures the essential mechanism by
which a nontrivial intermediate permutation can dominate. For replica number $N=3$, the spin variables take values in
\begin{equation}
  S_3=\{e,(12),(13),(23),(123),(132)\}.
\end{equation}
The Cayley distance table for $S_3$ is given in Table~\ref{tab:S3distancereview}.

\begin{table}[t]
\centering
\begin{tabular}{c|cccccc}
  $d(g,h)$ & $e$ & $(12)$ & $(13)$ & $(23)$ & $(123)$ & $(132)$ \\
  \hline
  $e$      & 0 & 1 & 1 & 1 & 2 & 2 \\
  $(12)$   & 1 & 0 & 2 & 2 & 1 & 1 \\
  $(13)$   & 1 & 2 & 0 & 2 & 1 & 1 \\
  $(23)$   & 1 & 2 & 2 & 0 & 1 & 1 \\
  $(123)$  & 2 & 1 & 1 & 1 & 0 & 2 \\
  $(132)$  & 2 & 1 & 1 & 1 & 2 & 0
\end{tabular}
\caption{Cayley distance table for $S_3$.}
\label{tab:S3distancereview}
\end{table}

Consider the boundary condition
\begin{equation}
  \{g_A,g_B,g_C\}=\{(123),(132),e\},
  \label{eq:negboundaryreview}
\end{equation}
which is gauge-equivalent to the standard negativity choice.
From Table~\ref{tab:S3distancereview}, one finds
\begin{equation}
  d(g_A,g_B)=d(g_B,g_C)=d(g_C,g_A)=2,
\end{equation}
so that
\begin{equation}
  S_{\rm pair}=6.
\end{equation}
If the bulk were restricted to the three phases
\begin{equation}
  \{e,(123),(132)\},
\end{equation}
then the natural configuration would be a symmetric $Y$-saddle with tension $2J$
on each wall.

However, the full group $S_3$ also contains the transpositions
\begin{equation}
  \tau\in\{(12),(13),(23)\},
\end{equation}
and for any such choice one has
\begin{equation}
  d(g_A,\tau)=d(g_B,\tau)=d(g_C,\tau)=1.
\end{equation}
Therefore
\begin{equation}
  S_\tau=3.
\end{equation}

In flat space, the criterion \eqref{eq:flatcriterionreview} becomes
\begin{equation}
  6 \ge \sqrt{3}\times 3,
\end{equation}
which is satisfied.
Equivalently,
\begin{equation}
  E_{(a)} \sim 6JL,
  \qquad
  E_{(b)} \sim 3\sqrt{3}JL,
\end{equation}
so the $\tau$-mediated saddle is energetically preferred.

In AdS, the leading comparison is instead
\begin{equation}
  6 = 2\times 3,
\end{equation}
so the near-boundary competition is saturated.
This is precisely the situation expected from the general bound
\eqref{eq:universalboundreview}.
At finite cutoff, the subleading terms lift this degeneracy and favor the
$\tau$-mediated saddle. 

Strictly speaking, logarithmic negativity is usually defined through an
even-replica continuation.
It is therefore useful to note that the same qualitative mechanism persists at
$N=4$.

For $N=4$, consider the boundary condition
\begin{equation}
  \{g_A,g_B,g_C\}=\{(1234),(1432),e\}.
  \label{eq:negboundaryreviewN4}
\end{equation}
These are the natural $N=4$ analogues of the $N=3$ data above.
From the Cayley distance table in Table~\ref{tab:S4-distance}, one finds
\begin{equation}
  d(g_A,g_B)=2,\qquad d(g_B,g_C)=3,\qquad d(g_C,g_A)=3,
\end{equation}
so that
\begin{equation}
  S_{\rm pair}=8.
\end{equation}
If the bulk were restricted to the three boundary-imposed phases alone, the
natural configuration would again be the naive $Y$-saddle.

However, Table~\ref{tab:S4-distance} also shows that the full group
$S_4$ contains nontrivial intermediate permutations that can lower the effective
tension. For example, if
\begin{equation}
  \tau=(12)(34),
\end{equation}
then one has
\begin{equation}
  d\bigl((1234),(12)(34)\bigr)=1,\qquad
  d\bigl((1432),(12)(34)\bigr)=1,\qquad
  d\bigl(e,(12)(34)\bigr)=2,
\end{equation}
so that
\begin{equation}
  S_\tau=4.
\end{equation}
In flat space, this gives
\begin{equation}
  8 \ge \sqrt{3}\times 4,
\end{equation}
which is satisfied.
In AdS, the leading comparison becomes
\begin{equation}
  8 = 2\times 4,
\end{equation}
so the criterion is again saturated.

Thus the same basic mechanism also survives at $N=4$: a nontrivial intermediate
permutation can compete with the naive $Y$-saddle. In this sense, the appearance
of such intermediate permutations is not an artifact of the $N=3$ case alone.
For the purposes of the present paper, however, the $N=3$ case captures the
essential domain-wall mechanism most transparently, while the $N=4$ case provides
a useful bridge to the $N=4$ multi-entropy analysis in the next section.



\section{Multi-entropy and 
the structural obstruction to replica symmetry breaking}
\label{sec:noRSB}

We now turn to the main structural result of this paper.
In contrast to the negativity example reviewed in the previous section, we will show that multi-entropy has a structural obstruction to replica symmetry breaking (RSB). 
This is not an accident of a few examples, but a structural consequence of the boundary permutations relevant to multi-entropy: they do not admit the kind of nontrivial intermediate permutation $\tau$ required for a dominant $\tau$-mediated saddle.

For the $\mathtt{q}$-partite R\'enyi multi-entropy with index $n$, the replica construction involves
\begin{equation}
  N=n^{\mathtt{q}-1}.
\end{equation}
The corresponding domain-wall spin variables therefore take values in $S_N$. In the tripartite case, the multi-entropy is specified by boundary twist insertions associated with three permutations $(g_A,g_B,g_C)$ acting on these replicas. We focus first on the case $\mathtt{q}=3$ and $n=2$, for which $N=4$ and the no-RSB statement can be established very explicitly. We then explain the structural reason behind this result and show how the same perspective extends to cases with higher $n$ and higher $q$.

\subsection{The $n=2$ multi-entropy}

For the $n=2$ multi-entropy, the relevant replica number is $N=4$, so the spin variables take values in $S_4$. The corresponding boundary permutations follow from the replica construction of multi-entropy~\cite{Gadde:2022cqi,Penington:2022dhr,Gadde:2023zzj}. Up to gauge equivalence, the tripartite boundary data may be chosen as
\begin{equation}
  g_A=(12)(34),\qquad
  g_B=(13)(24),\qquad
  g_C=e.
  \label{eq:n2ME_boundary}
\end{equation}

From the Cayley distance table for $S_4$ in Table~\ref{tab:S4-distance}, one finds
\begin{equation}
  d(g_A,g_B)=d(g_B,g_C)=d(g_C,g_A)=2,
\end{equation}
so that
\begin{equation}
  S_{\rm pair}
  :=
  d(g_A,g_B)+d(g_B,g_C)+d(g_C,g_A)
  =
  6.
  \label{eq:n2ME_Spair}
\end{equation}

Now consider a possible intermediate permutation
\begin{equation}
  \tau \notin \{g_A,g_B,g_C\},
\end{equation}
and define
\begin{equation}
  S_\tau:=d(g_A,\tau)+d(g_B,\tau)+d(g_C,\tau).
\end{equation}
A direct inspection of the relevant $S_4$ distances shows that for every such $\tau$,
\begin{equation}
  S_\tau \ge 5.
  \label{eq:n2ME_Stau_lower}
\end{equation}

This is already enough to exclude a dominant $\tau$-mediated saddle.
Indeed, in flat space one would need
\begin{equation}
  S_{\rm pair}\ge \sqrt{3}\,S_\tau,
\end{equation}
but \eqref{eq:n2ME_Spair} and \eqref{eq:n2ME_Stau_lower} give
\begin{equation}
  6 < \sqrt{3}\times 5.
\end{equation}
In AdS, the corresponding condition is
\begin{equation}
  S_{\rm pair}\ge 2S_\tau,
\end{equation}
which is even more strongly violated:
\begin{equation}
  6 < 2\times 5 = 10.
\end{equation}
Thus the $n=2$ multi-entropy does not exhibit RSB.

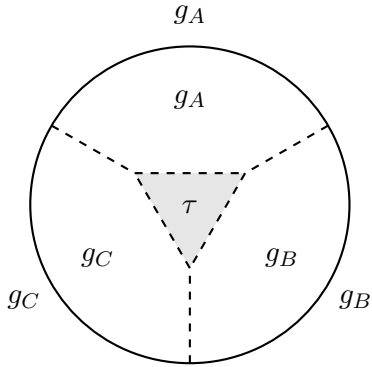
\begin{figure}[t]
  \centering
  \makebox[\linewidth][c]{%
    \begin{tikzpicture}[scale=1.4]
      \draw[thick] (0,0) circle (1.5);

      \coordinate (A) at (150:1.5);
      \coordinate (B) at (30:1.5);
      \coordinate (C) at (270:1.5);

      \coordinate (a) at (150:0.60);
      \coordinate (b) at (30:0.60);
      \coordinate (c) at (270:0.60);

      \fill[gray!20] (a) -- (b) -- (c) -- cycle;

      \draw[thick, dashed] (A) -- (a);
      \draw[thick, dashed] (B) -- (b);
      \draw[thick, dashed] (C) -- (c);

      \draw[thick, dashed] (a) -- (b);
      \draw[thick, dashed] (b) -- (c);
      \draw[thick, dashed] (c) -- (a);

      \node at (90:1.8)   {$g_A$};
      \node at (330:1.8)  {$g_B$};
      \node at (210:1.8)  {$g_C$};

      \node at (90:1.0)   {$g_A$};
      \node at (330:1.0)  {$g_B$};
      \node at (210:1.0)  {$g_C$};

      \node at (0,-0.02) {$\tau$};
    \end{tikzpicture}%
  }
  \caption{A triangular $\tau$-bubble configuration with the same boundary anchoring as the naive $Y$-configuration. The central $\tau$ domain is taken to be an equilateral triangle, representing the local flat-space approximation to a small interior $\tau$-bubble.}
  \label{fig:tau-triangle-bubble}
\end{figure}

It is worth emphasizing what is excluded by the above estimate.
The AdS argument above showing that the $n=2$ multi-entropy does not exhibit RSB implies that the configuration in Fig.~\ref{fig:tau-mediated-curved} never dominates over the naive $Y$-configuration in Fig.~\ref{fig:naive-Ygeneric}. Furthermore, the flat-space argument above also rules out a bulk triangular $\tau$-bubble nucleated near the bulk interior, as in Fig.~\ref{fig:tau-triangle-bubble}. This is because the geometry is locally approximately flat in such a small region, so the same $\sqrt{3}$ cost applies, and the estimate above again shows that this configuration cannot dominate over the naive $Y$-configuration.

\subsection{The $n=3$ multi-entropy: another explicit low-replica check}

It is useful to complement the $n=2$ analysis with the next nontrivial tripartite case,
namely $n=3$. In this case, the replica number is $N=9$,
and, up to gauge equivalence, the tripartite boundary permutations may be chosen as
\begin{equation}
g_A=(123)(456)(789),\qquad
g_B=(147)(258)(369),\qquad
g_C=e.
\end{equation}
These are the natural $n=3$ analogues of the row- and column-cycle permutations discussed
more generally below.

From the cycle structure, one immediately finds
\begin{equation}
d(g_A,g_B)=d(g_B,g_C)=d(g_C,g_A)=6,
\end{equation}
so that
\begin{equation}
S_{\rm pair}:=d(g_A,g_B)+d(g_B,g_C)+d(g_C,g_A)=18.
\end{equation}
Now consider a possible intermediate permutation
\begin{equation}
\tau\notin \{g_A,g_B,g_C\},
\end{equation}
and define, as before,
\begin{equation}
S_\tau:=d(g_A,\tau)+d(g_B,\tau)+d(g_C,\tau).
\end{equation}
By an exhaustive numerical check over all $9!$ elements of $S_9$, one finds that
\begin{equation}
S_\tau \ge 12
\end{equation}
for all $\tau \in S_9$.

This is already enough to exclude a $\tau$-mediated configuration of Fig.~\ref{fig:tau-mediated-curved} type in AdS, since
\begin{equation}
18 < 2 \times 12
\qquad \Rightarrow \qquad
S_{\rm pair} < 2S_\tau \, .
\end{equation}
Similarly, a triangular $\tau$-bubble configuration of Fig.~\ref{fig:tau-triangle-bubble} type is also excluded, since
\begin{equation}
18 < \sqrt{3}\times 12
\qquad \Rightarrow \qquad
S_{\rm pair} < \sqrt{3}\,S_\tau \, .
\end{equation}
Thus, even in the $n=3$ tripartite multi-entropy case, both the $\tau$-mediated AdS configuration and the triangular $\tau$-bubble configuration are energetically disfavored relative to the naive $Y$-configuration.

This $n=3$ example is useful because it shows that the simple no-go mechanism seen at $n=2$ persists at the next replica number as well.
It is natural to ask whether this can be extended further. A complete check quickly becomes impractical: already for $n=4$, one would have to deal with $16! = \order{10^{15}}$ permutations. Nevertheless, one can find explicit nontrivial $\tau$ examples with $S_\tau=22$, while $S_{\rm pair}=36$. Since $36 < 22 \sqrt{3}$,  even such candidates for $\tau$ remain above the threshold for the symmetric triangular $\tau$-bubble to dominate. We leave a systematic analysis of higher $n$ for future work.

Before closing this subsection, let us comment on an important difference between RTNs and a single random tensor model.
For a single random tensor model, finding a nontrivial $\tau$ such that
\begin{equation}
S_\tau  <  d(g_A,r)+d(g_B,r)+d(g_C,r)
\qquad \mbox{where } r \in \{g_A,g_B,g_C\},
\end{equation}
might already be enough to indicate RSB, and in fact the above $\tau$ satisfies this inequality because $d(g_A,e)+d(g_B,e)=24$. However, in random tensor networks, this is not enough. As we have shown, there is an additional geometrical factor that must be taken into account\footnote{We thank Simon Lin for asking about this point.}.

In the next subsections, we explain the structural reason behind the AdS no-go result.


\subsection{A universal upper bound and geodesic saturation in AdS}

We now formulate the AdS criterion for a $\tau$-mediated saddle to compete with the naive tripartite saddle. The point of this subsection is that the argument is completely general for tripartite boundary data: it does not rely on the special form of the $n=2$ multi-entropy boundary conditions discussed in the previous subsection.

For a tripartite observable with boundary data $(g_A,g_B,g_C)$ and an intermediate candidate $\tau$, define
\begin{align}
  S_{\rm pair} &:= d(g_A,g_B)+d(g_B,g_C)+d(g_C,g_A), \\
  S_\tau &:= d(g_A,\tau)+d(g_B,\tau)+d(g_C,\tau).
\end{align}
As reviewed in Sec.~2, in AdS a $\tau$-mediated saddle can compete with the naive saddle only if
\begin{equation}
  S_{\rm pair} \ge 2S_\tau.
  \label{eq:AdS_RSB_condition_sec3}
\end{equation}
This is the general AdS version of the RSB criterion.

On the other hand, the Cayley distance is a genuine metric and therefore obeys the triangle inequality. Applying it to the three pairs $(g_A,g_B)$, $(g_B,g_C)$, and $(g_C,g_A)$ with intermediate point $\tau$, one obtains
\begin{align}
  d(g_A,g_B) &\le d(g_A,\tau)+d(\tau,g_B), \nonumber\\
  d(g_B,g_C) &\le d(g_B,\tau)+d(\tau,g_C), \nonumber\\
  d(g_C,g_A) &\le d(g_C,\tau)+d(\tau,g_A). \nonumber
\end{align}
Summing these inequalities yields the universal upper bound
\begin{equation}
  S_{\rm pair} \le 2S_\tau.
  \label{eq:universal_upper_bound_sec3}
\end{equation}
Therefore the AdS condition \eqref{eq:AdS_RSB_condition_sec3} can be satisfied only if it is saturated:
\begin{equation}
  S_{\rm pair}=2S_\tau.
  \label{eq:saturation_sec3}
\end{equation}

Equivalently, each of the triangle inequalities above must saturate separately, so that $\tau$ must lie on geodesics between all three relevant pairs of boundary permutations:
\begin{equation}
  d(g_i,g_j)=d(g_i,\tau)+d(\tau,g_j),
  \qquad
  g_i,g_j\in\{g_A,g_B,g_C\}.
  \label{eq:geodesic_saturation_sec3}
\end{equation}
Thus, in AdS the search for RSB reduces to the existence of a nontrivial common geodesic intermediate permutation $\tau$. In a gauge where one boundary value is set to the identity, the condition becomes that $\tau$ must lie simultaneously on geodesics from $e$ to each of the remaining boundary permutations. For tripartite boundary data, this is equivalent to asking whether there exists a nontrivial $\tau$ lying on both a geodesic from $e$ to $g_A$ and a geodesic from $e$ to $g_B$.

This reformulation is important because it turns the no-RSB question into a structural one. To exclude RSB, it is enough to show that no nontrivial $\tau$ can satisfy \eqref{eq:geodesic_saturation_sec3} for the boundary permutations relevant to multi-entropy. In this sense, the question is no longer merely whether a few numerical candidates fail, but whether the boundary data themselves admit a common intermediate permutation at all.

The following subsections make this obstruction explicit. We first reinterpret the $n=2$ multi-entropy discussed above from this geodesic-saturation viewpoint, and then use the same perspective as a guide for higher-replica cases.

\subsection{The structural obstruction for tripartite multi-entropy}

We now make the geodesic-saturation criterion explicit for tripartite multi-entropy. The basic lesson is that the failure of RSB is not accidental: the boundary permutations do not admit a nontrivial common geodesic intermediate permutation $\tau$ satisfying \eqref{eq:geodesic_saturation_sec3}.

\subsubsection*{$n=2$ as a warm-up}

For the $n=2$ multi-entropy, the boundary permutations are
\begin{equation}
  g_A=(12)(34),\qquad
  g_B=(13)(24),\qquad
  g_C=e.
\end{equation}
The explicit check in Sec.~3.1 already showed that no $\tau$-mediated saddle can dominate. We now reinterpret this from the geodesic-saturation viewpoint.

Since
\begin{equation}
  d(e,g_A)=2,
\end{equation}
a necessary condition for \eqref{eq:geodesic_saturation_sec3} to hold between $e$ and $g_A$ is that $\tau$ lie on a geodesic from $e$ to $g_A$.

For the Cayley metric generated by transpositions, a geodesic from $e$ to
\begin{equation}
  g_A=(12)(34)
\end{equation}
is obtained by successively introducing the transpositions appearing in $g_A$. Therefore any intermediate point on such a geodesic must be built from a subset of these transpositions. In particular, if
\begin{equation}
  d(e,\tau)=1,
\end{equation}
then necessarily
\begin{equation}
  \tau\in\{(12),(34)\}.
\end{equation}
Similarly, if
\begin{equation}
  d(e,\tau)=2,
\end{equation}
then the only geodesic endpoint is
\begin{equation}
  \tau=(12)(34)=g_A,
\end{equation}
which is excluded for a genuine RSB saddle. Thus the only nontrivial candidates compatible with the pair $(e,g_A)$ are
\begin{equation}
  \tau=(12)\qquad\text{or}\qquad\tau=(34).
\end{equation}

However, for either of these choices, the corresponding relation with
\begin{equation}
  g_B=(13)(24)
\end{equation}
fails. Indeed, one finds
\begin{equation}
  d(e,g_B)=2,\qquad
  d(e,\tau)=1,\qquad
  d(\tau,g_B)=3,
\end{equation}
so that
\begin{equation}
  d(e,g_B)\neq d(e,\tau)+d(\tau,g_B).
\end{equation}
Hence the geodesic-saturation condition cannot be satisfied simultaneously for the pairs $(e,g_A)$ and $(e,g_B)$.

The structural reason is simple: the transpositions appearing in $g_A$ are not shared with those appearing in $g_B$. Therefore a permutation $\tau$ that lies on a geodesic from $e$ to $g_A$ cannot at the same time lie on a geodesic from $e$ to $g_B$. In other words, the boundary data do not possess a common overlapping transposition structure.

\subsubsection*{General tripartite case}

We now show that the same obstruction persists for arbitrary R\'enyi index $n$ in the tripartite case.

For tripartite multi-entropy, the replica number is
\begin{equation}
  N=n^2.
\end{equation}
Up to gauge equivalence, the boundary permutations may be chosen as
\begin{align}
g_A&=(1,2,\dots,n)(n+1,n+2,\dots,2n)\cdots(n^2-n+1,n^2-n+2,\dots,n^2),\\
g_B&=(1,n+1,\dots,n^2-n+1)(2,n+2,\dots,n^2-n+2)\cdots(n,2n,\dots,n^2),\\
g_C&=(1)(2)\cdots(n^2)=e.
\end{align}
Thus $g_A$ is a product of $n$ disjoint $n$-cycles acting along the ``row'' blocks, while $g_B$ is a product of $n$ disjoint $n$-cycles acting along the ``column'' blocks.  

It is convenient to define these blocks explicitly. For $a=0,1,\dots,n-1$, let
\begin{equation}
  R_a:=\{an+1,an+2,\dots,an+n\},
\end{equation}
and for $b=1,2,\dots,n$, let
\begin{equation}
  C_b:=\{b,b+n,b+2n,\dots,b+(n-1)n\}.
\end{equation}
Then the cycles of $g_A$ are precisely the row blocks $R_a$, while the cycles of $g_B$ are precisely the column blocks $C_b$.

Suppose now that a nontrivial permutation $\tau$ satisfies the geodesic-saturation condition \eqref{eq:geodesic_saturation_sec3}. Since $g_C=e$, this requires in particular that $\tau$ lie simultaneously on a geodesic from $e$ to $g_A$ and on a geodesic from $e$ to $g_B$.

We first analyze the condition that $\tau$ lie on a geodesic from $e$ to $g_A$. Since
\begin{equation}
  d(e,g_A)=N-C(g_A)=n^2-n,
\end{equation}
any geodesic from $e$ to $g_A$ has length $n^2-n$. Along such a geodesic, each multiplication by a transposition reduces the number of cycles by exactly one. Thus, starting from the identity permutation, the cycles are built only by successive mergers of previously disjoint cycles; they are never split along the geodesic.

Because the final permutation $g_A$ has cycles exactly equal to the row blocks $R_a$, no geodesic from $e$ to $g_A$ can ever merge elements belonging to different row blocks. If such a merger occurred at some intermediate stage, the resulting mixed cycle could only be undone by a later splitting operation, but a geodesic from $e$ to $g_A$ never increases the number of cycles. Therefore every cycle of an intermediate permutation $\tau$ on a geodesic from $e$ to $g_A$ must be contained entirely within a single row block $R_a$.

By the same argument, if $\tau$ lies on a geodesic from $e$ to $g_B$, then every cycle of $\tau$ must be contained entirely within a single column block $C_b$.

We now combine the two conditions. If $\tau$ lies simultaneously on a geodesic from $e$ to $g_A$ and on a geodesic from $e$ to $g_B$, then every cycle of $\tau$ must be contained both in some row block $R_a$ and in some column block $C_b$. But
\begin{equation}
  R_a\cap C_b
\end{equation}
contains exactly one element. Hence every cycle of $\tau$ must be a singleton, which implies
\begin{equation}
  \tau=e.
\end{equation}

Therefore there is no nontrivial common geodesic intermediate permutation $\tau$ for the tripartite multi-entropy boundary data. 
By the criterion of \eqref{eq:geodesic_saturation_sec3}, 
this means that the tripartite multi-entropy has a structural obstruction to replica symmetry breaking for any R\'enyi index $n$.

This general tripartite argument clarifies the basic mechanism already visible in the $n=2$ warm-up example. The obstruction is not numerical but structural: the row-cycle structure of $g_A$ and the column-cycle structure of $g_B$ are incompatible with the existence of a nontrivial common intermediate permutation.

In the next subsection, we extend this perspective from the tripartite case to higher multi-partite number $q$.

\subsection{Extension to higher $\mathtt{q} \ge 4$-partite multi-entropy cases}

The obstruction found above is not special to the lowest-replica examples.
What matters is not the detailed value of $n$ or $q$ by itself, but the structural form of the boundary permutations associated with multi-entropy.
We now explain this general mechanism\footnote{After completion of this work, we were informed of \cite{Carrozza:2026qcf}, where a related structural incompatibility of the permutations relevant to multi-entropies is also discussed.}.

For the $\mathtt{q}$-partite multi-entropy with R\'enyi index $n$, the replica labels may be organized as points on a $(\mathtt{q}-1)$-dimensional discrete lattice of size
\begin{equation}
  n\times n\times\cdots\times n,
\end{equation}
with total number of points
\begin{equation}
  N=n^{\mathtt{q}-1}.
\end{equation}
It is convenient to label each replica by a multi-index
\begin{equation}
  (a_1,a_2,\dots,a_{\mathtt{q}-1}),
  \qquad
  a_k\in\mathbb{Z}_n.
\end{equation}
In this language, each boundary permutation $g_i$ acts cyclically along one replica direction while keeping the others fixed.
Thus the boundary permutations define natural block decompositions of the replica labels into disjoint $n$-element subsets aligned along different coordinate directions of this replica hypercube.

More concretely, for the next nontrivial quadri-partite case $\mathtt{q}=4$, the replica labels may be organized as points on a three-dimensional $n\times n\times n$ lattice, so that
\begin{equation}
  N=n^3.
\end{equation}
Writing the replica labels as triples
\begin{equation}
  (a,b,c)\in \mathbb{Z}_n^3,
\end{equation}
the boundary permutations may be chosen so that
\begin{align}
  g_A &: (a,b,c)\mapsto (a+1,b,c), \\
  g_B &: (a,b,c)\mapsto (a,b+1,c), \\
  g_C &: (a,b,c)\mapsto (a,b,c+1), \\
  g_D &: (a,b,c)\mapsto (a,b,c),
\end{align}
with addition understood mod $n$. Thus $g_A$, $g_B$, and $g_C$ act cyclically along the three coordinate directions of the replica hypercube, while $g_D=e$.

Now suppose that a permutation $\tau$ satisfies the geodesic-saturation condition
\begin{equation}
  d(e,g_i)=d(e,\tau)+d(\tau,g_i)
  \label{eq:general_geodesic_condition_q}
\end{equation}
for a given boundary permutation $g_i$. Then $\tau$ must lie on a geodesic from $e$ to $g_i$.
As in the low-replica examples, this implies that $\tau$ cannot mix labels belonging to different final cycle supports of $g_i$.
Otherwise such mixing would have to be undone later in order to recover the final cycle structure of $g_i$, which is impossible along a geodesic.
Hence $\tau$ must preserve the block decomposition associated with $g_i$.

For RSB in AdS, however, $\tau$ must satisfy this condition simultaneously for all relevant boundary permutations.
Thus a nontrivial common $\tau$ can exist only if the block decompositions associated with the different replica directions admit a nontrivial common refinement.
But for multi-entropy these block decompositions are precisely the coordinate-aligned decompositions of the replica hypercube, and they are mutually incompatible.
A nontrivial permutation that preserves the blocks along one coordinate direction necessarily conflicts with the blocks defined by another direction.
Therefore the only permutation preserving all such decompositions simultaneously is the identity:
\begin{equation}
  \tau=e.
\end{equation}
But $\tau=e$ is not a genuine intermediate permutation and does not define a nontrivial $\tau$-mediated saddle.
Hence there is no genuine RSB saddle of Fig.~\ref{fig:tau-mediated-curved} type.

This becomes especially transparent already in the first nontrivial higher-partite example, namely $q=4$ and $n=2$.
In that case the replica labels may be arranged as the vertices of a $2\times 2\times 2$ cube, so $N=8$.
Up to gauge equivalence, one may choose
\begin{align}
  g_A&=(12)(34)(56)(78),\qquad  g_B=(13)(24)(57)(68),\nonumber \\
  g_C&=(15)(26)(37)(48), \qquad   g_D=e. 
\end{align}
These correspond to pairings along the three coordinate directions of the cube, namely the $x$-, $y$-, and $z$-directions, respectively.

Suppose that a permutation $\tau$ satisfies the geodesic-saturation condition with respect to the pairs $(e,g_A)$, $(e,g_B)$, and $(e,g_C)$.
Then, by the same reasoning as in the tripartite case, each cycle of $\tau$ must be contained in one of the $x$-direction pairs, and likewise in one of the $y$-direction pairs and one of the $z$-direction pairs.
But the intersection of an $x$-pair, a $y$-pair, and a $z$-pair consists of a single vertex.
Hence every cycle of $\tau$ must be a singleton, so $\tau=e$.
Therefore there is no nontrivial common intermediate permutation.

In this way, the absence of AdS RSB saddle of Fig.~\ref{fig:tau-mediated-curved} type for multi-entropy is a structural consequence of the replica geometry of the boundary data.
The issue is not that a few candidates happen to fail, but that the boundary permutations themselves do not admit a nontrivial common geodesic intermediate permutation.

\subsection{Comparison with negativity}

It is useful to summarize the contrast with negativity in structural terms.

For negativity, the boundary permutations do admit a nontrivial common intermediate structure.
In the $N=3$ example reviewed in Sec.~2, the three boundary permutations
\begin{equation}
  \{(123),(132),e\}
\end{equation}
all lie at distance one from a transposition $\tau$.
Equivalently, there exists a nontrivial permutation $\tau$ such that
\begin{equation}
  d(g_A,\tau)=d(g_B,\tau)=d(g_C,\tau)=1,
\end{equation}
so that
\begin{equation}
  S_\tau=3,
  \qquad
  S_{\rm pair}=6,
\end{equation}
and hence
\begin{equation}
  S_{\rm pair}=2S_\tau.
\end{equation}
Thus the AdS geodesic-saturation condition is satisfied, and negativity 
admits a nontrivial common geodesic mediator.

For multi-entropy, by contrast, the relevant boundary permutations are organized very differently.
Although each boundary permutation can individually be connected to the identity by a geodesic, there is no single nontrivial $\tau$ that lies on all the relevant geodesics simultaneously.
In the $n=2$ case, this failure appears as the incompatibility of the transposition supports.
In the tripartite case with general $n$, it appears as the incompatibility of the block decompositions defined by the disjoint $n$-cycle supports.
More generally, for higher multi-partite multi-entropy, the same obstruction arises from the mutually incompatible coordinate-aligned block decompositions of the replica hypercube.

Therefore the distinction is not quantitative but structural.
Negativity admits a nontrivial common intermediate permutation $\tau$ satisfying the AdS geodesic-saturation condition, whereas multi-entropy does not.
This is the essential structural difference between negativity and multi-entropy in the original domain-wall spin model.

\subsection{General lesson for multi-entropy}

We can now summarize the main lesson of this section.

In the RTN domain-wall spin model, multi-entropy behaves qualitatively differently from negativity.
For negativity, the boundary permutations admit a nontrivial intermediate permutation $\tau$ that is sufficiently close to all boundary sectors.
Equivalently, in AdS there exists a nontrivial $\tau$ satisfying the geodesic-saturation condition, and in flat space the corresponding $\tau$-mediated saddle can lower the domain-wall energy.
This is why negativity exhibits replica symmetry breaking.

For multi-entropy, by contrast, the boundary permutations do not possess the required common intermediate structure.
In the $n=2$ case, this appears as the absence of common transposition support.
In the tripartite case with general $n$, it appears as the incompatibility of the relevant cycle-support block decompositions.
In higher multi-partite cases, the same mechanism is naturally described in terms of incompatible coordinate-direction decompositions of the replica hypercube.
In all cases, the consequence is the same: no genuinely nontrivial permutation $\tau$ can simultaneously satisfy the geodesic-saturation condition for all relevant boundary sectors.

We therefore arrive at the following conclusion for the RTN domain-wall spin model:
\begin{quote}
\it
In contrast to negativity, multi-entropy has a structural obstruction to replica symmetry breaking in the RTN domain-wall spin model.
\rm
\end{quote} 

This concludes the ungauged analysis.
In the next section, we ask whether this conclusion can be modified once a minimal gauge structure is introduced.

\section{A toy gauge extension} 
\label{sec:gauge}

\subsection{Motivation for a gauge extension and the set-up}

In the previous section, we showed that multi-entropy does not exhibit bulk replica symmetry breaking within the conventional RTN domain-wall spin model.
A natural next question is whether this no-RSB property remains robust once one moves beyond this simplest setting.

This question is worth asking here because the present section is intended as a first probe beyond the simplest RTN domain-wall framework.
That framework already captures an important part of the fixed-area, leading-order structure of gravitational entanglement~\cite{Dong:2018seb}, but it is natural to ask whether the replica structure changes once one incorporates ingredients such as gauge redundancy and global constraints.
A related motivation is that, in the language of holographic error correction, the conventional RTN/domain-wall framework has trivial area operators~\cite{Harlow:2016vwg}; incorporating a gauge field gives rise to nontrivial area operators \cite{Donnelly:2016qqt, Akers:2018fow, Dong:2023kyr, Akers:2024ixq}.

In this section, we consider a simple gauge extension of the RTN domain-wall spin model, following the general perspective of~\cite{Akers:2024ixq} (see also~\cite{Akers:2024wab,Dong:2023kyr,Qi:2022lbd,Donnelly:2016qqt}).
Rather than attempting to capture the full gravitational problem, we use this toy construction to ask whether introducing gauge structure can modify the bulk replica pattern seen in the conventional RTN domain-wall spin-model framework.

A key structural feature of conventional random tensor networks is that internal edges are contracted using maximally entangled Bell states. In the large $D$ limit, this leads to the usual domain-wall description, in which an edge separating permutation sectors $\sigma$ and $\pi$ contributes a factor $D^{-d(\sigma,\pi)}$. As a result, the conventional RTN domain-wall spin-model construction provides both a local bulk description and the usual Ryu--Takayanagi-type minimal-cut picture. A brief review of this standard RTN construction is given in Appendix~\ref{sec:app-crtn}.

In the present section, we replace this Bell-pair structure by a gauge-invariant bulk state obtained from a lattice gauge theory.
We focus on the simple case
\begin{equation}
  G=\mathbb Z_2,
\end{equation}
for which the gauge-invariant states become stabilizer states and explicit numerical analysis becomes possible using the techniques of~\cite{Akella:2026xza}.
Further details of the lattice-gauge-theory ingredients and of the construction are collected in Appendix~\ref{sec:app-lgt}.

In the following subsection, we explain the proposal in more detail and then present numerical evidence for the resulting bulk replica structure in this toy $\mathbb Z_2$ gauge extension.

\subsection{A proposal for a gauge-invariant state replacing Bell pairs}

We begin with the state $\ket{0 \dots 0}$ in the pre-gauged Hilbert space, where each edge variable is in the identity element $0$ of the gauge group $G$. 
Here we use the notation reviewed in Appendix~\ref{sec:app-lgt}.
Imposing the gauge constraints then prepares the following gauge-invariant state: 
\begin{equation}
 \ket{\mathcal{E}} = \left(\bigotimes_{v \in V_{bulk}} A_v \right)\ket{0 \dots 0}.
\end{equation}
This is a state of the gauge degrees of freedom on all internal edges.

We now construct a gauge-invariant state that replaces the Bell-pair contractions in the RTN construction. However, the Bell state on an edge lives on two copies of the edge Hilbert space. We therefore use the \emph{factorization map} in lattice gauge theory. Recall that the factorization map $S: L^2(G) \to L^2(G) \otimes L^2(G)$ is defined by
\begin{equation}
 S \ket{g} = \frac{1}{\sqrt{|G|}} \sum_{h \in G} \ket{g h^{-1}} \otimes \ket{h},
\end{equation}
and maps one copy of the edge Hilbert space to two copies. We then apply this map to each edge and use the resulting state for the RTN contraction in place of the Bell pair.

Applying the factorization map to the reference state $\ket{0}$ naturally produces a maximally entangled Bell pair. Therefore, if one omits the gauge constraints from the first step, the present construction reduces to the conventional random tensor network.

Let us consider the simple example of a single edge connecting two vertices in a $\mathbb{Z}_2$ gauge theory,  where our proposal goes as follows.
First, we start with the state $\ket{0}$ in the pre-gauged Hilbert space.
The gauge constraint at each vertex is imposed by the projector $P=(I+X)/2$, which maps $\ket{0}$ to the unnormalized state $(\ket{0}+\ket{1})/2 \propto \ket{+}$.
After normalizing this state, we apply the factorization map, which gives $\ket{++} = (\ket{00}+\ket{11}+\ket{10}+\ket{01})/2$.
Here, the labels $0$ and $1$ denote the $\mathbb{Z}_2$ variables living on the edges, {\it not} the permutation-valued spin variables introduced in the RTN domain-wall spin model. Note also that this one-edge example is only meant to illustrate how the gauge projection and the factorization map work in the simplest setting.

For a more nontrivial graph, the same construction gives a nontrivial state on the edges. In the case of $\mathbb{Z}_2$ gauge theory, the resulting state is a stabilizer state \cite{Gottesman:1997zz}. When we use this state instead of the maximally entangled Bell state in the conventional RTN construction, we obtain a sum over multi-invariants of stabilizer states. We use the numerical technique developed in \cite{Akella:2026xza} to calculate such multi-invariants. 
In the next subsection, we present the results of our numerical analysis.

\subsection{Numerical results}

\subsubsection{Star graph: the depth-1 model}

\begin{figure}
 \centering
 \includegraphics[width=0.3\textwidth]{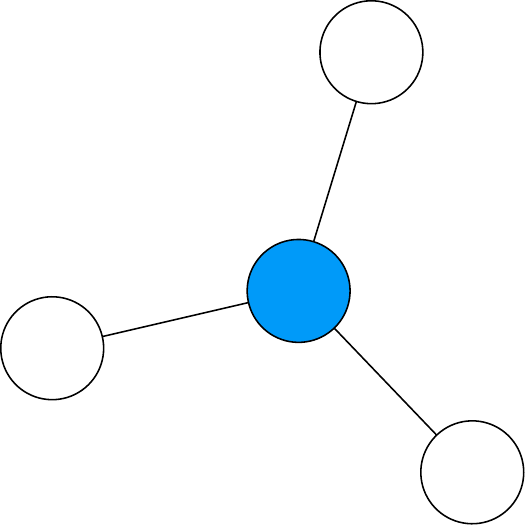}
 \caption{A star graph with one central bulk vertex (solid) and three boundary vertices (hollow).}
 \label{fig:star}
\end{figure}

As the simplest example, let us consider the graph shown in Fig. \ref{fig:star}. The central solid vertex is where we impose the $\mathbb{Z}_2$-gauge constraints. 
We begin with the reference state in which all three edge variables are in the identity state, namely $\ket{000}$, where the labels $0$ and $1$ refer to the $\mathbb{Z}_2$ variables on the edges. The gauge projection at the central vertex then correlates the three edges by producing the superposition of the configuration $\ket{000}$ and the configuration $\ket{111}$. In this way, before applying the factorization map, the gauge projection prepares a 3-qubit GHZ state on the three edges.

We next apply the factorization map to each of the three edges. Under this map, the edge state $\ket{0}$ is sent to the Bell state $\ket{\tilde{0}}$, while the edge state $\ket{1}$ is sent to the orthogonal Bell state $\ket{\tilde{1}}$, where
\begin{equation}
 \ket{\tilde{0}} = \frac{1}{\sqrt{2}}\ket{00} + \frac{1}{\sqrt{2}}\ket{11}, \qquad
 \ket{\tilde{1}} = \frac{1}{\sqrt{2}}\ket{01} + \frac{1}{\sqrt{2}}\ket{10}.
\end{equation}
As a result, the 3-qubit GHZ state on the three edges is mapped to the following 6-qubit state:
\begin{equation}\label{eq:gi-state}
 \ket{\Psi} = \frac{1}{\sqrt{2}}\ket{\tilde{0}\tilde{0}\tilde{0}} + \frac{1}{\sqrt{2}}\ket{\tilde{1}\tilde{1}\tilde{1}}.
\end{equation}

We then assign three permutations $g_A$, $g_B$, and $g_C$ at the three boundary vertices and ask what permutation $\tau$ at the central vertex minimizes the energy. Note that, if one omits the gauge constraints, the state $\ket{\Psi}$ in \eqref{eq:gi-state} reduces to the product state of the Bell state, $\ket{\tilde{0}\tilde{0}\tilde{0}}$. In that case, as we have seen in Sec.~\ref{sec:noRSB}, the problem reduces to finding the permutation $\tau$ such that
$d(g_A, \tau) + d(g_B, \tau) + d(g_C, \tau)$
is minimized. 

With the gauge-invariant state $\ket{\Psi}$ in \eqref{eq:gi-state}, however, this is no longer the case. This is because the gauge-invariant state $\ket{\Psi}$ is not local across individual edges. We therefore numerically evaluate the corresponding contribution with the gauge-invariant state and minimize over all permutations $\tau$. With this setup in place, we can now turn to the numerical results.

\begin{figure}
 \centering
 \includegraphics[width = 0.4\textwidth]{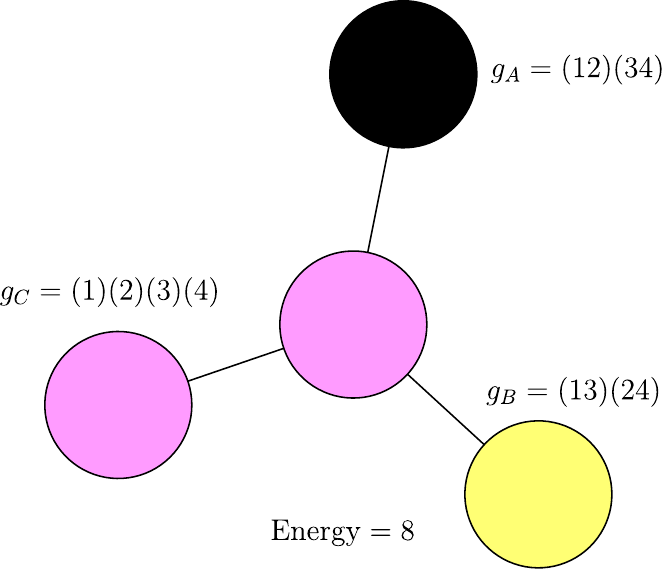}
 \hfill
 \includegraphics[width = 0.45\textwidth]{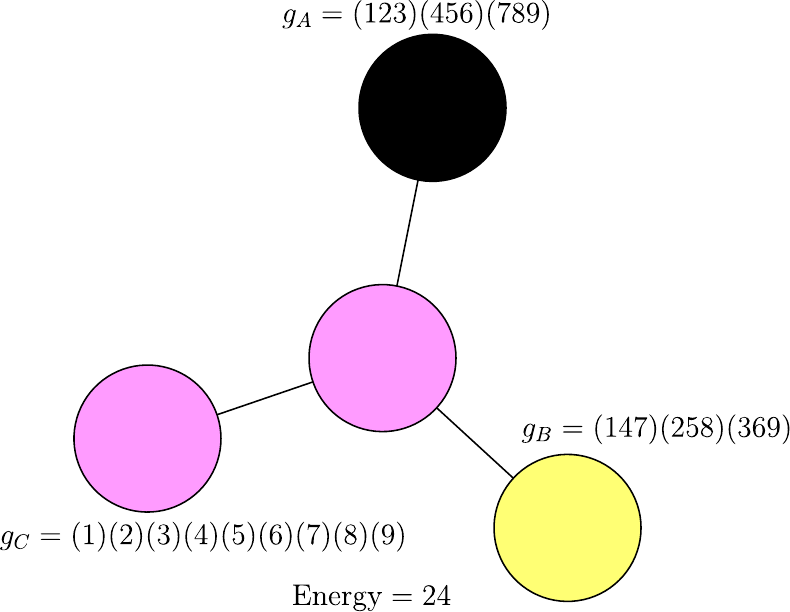}
 \caption{The minimum bulk configurations on the star graph for the $n = 2$ and the $n = 3$ R\'enyi multi-entropy. The central bulk vertex carries the permutation $\tau = g_A$ thereby preserving replica symmetry. } 
 \label{fig:star-n-multi}
\end{figure}

Our numerical minimization shows that if we pick $(g_A, g_B, g_C)$ to be the ones that define the $n = 2$ R\'enyi multi-entropy, {\it i.e.}, $g_A = (12)(34)$, $g_B = (13)(24)$, and $g_C = (1)(2)(3)(4)$, then the minimal $\tau$ belongs to  $\{g_A, g_B, g_C\}$. In other words, replica symmetry is preserved. Similarly, when we pick $(g_A, g_B, g_C)$ to be the ones that define the $n = 3$ R\'enyi multi-entropy, {\it i.e.}, $g_A = (123)(456)(789)$, $g_B = (147)(258)(369)$ and $g_C = (1)(2)(3)(4)(5)(6)(7)(8)(9)$,  then the minimal $\tau$ can be chosen to belong to $\{g_A, g_B, g_C\}$. Once again, replica symmetry is preserved. See Fig. \ref{fig:star-n-multi} for an illustration.

\begin{figure}
 \centering
 \includegraphics[width=0.45\textwidth]{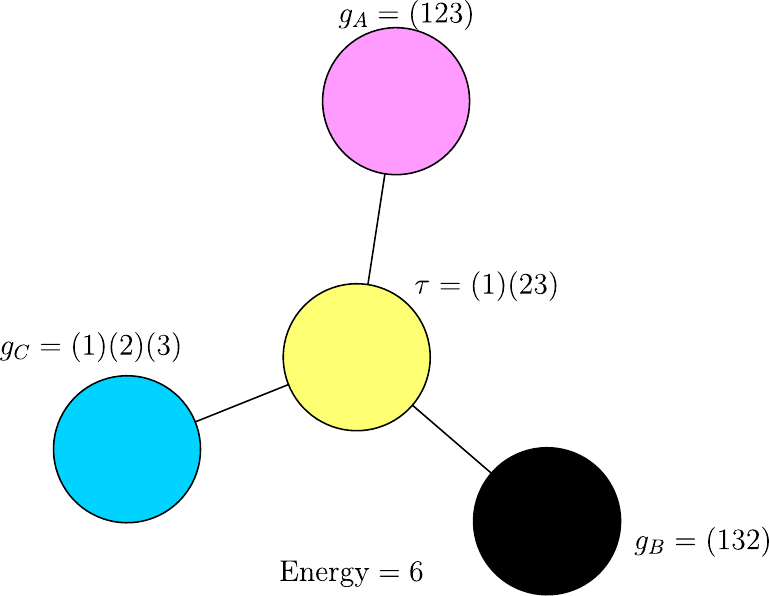}
 \hfill
 \includegraphics[width=0.45\textwidth]{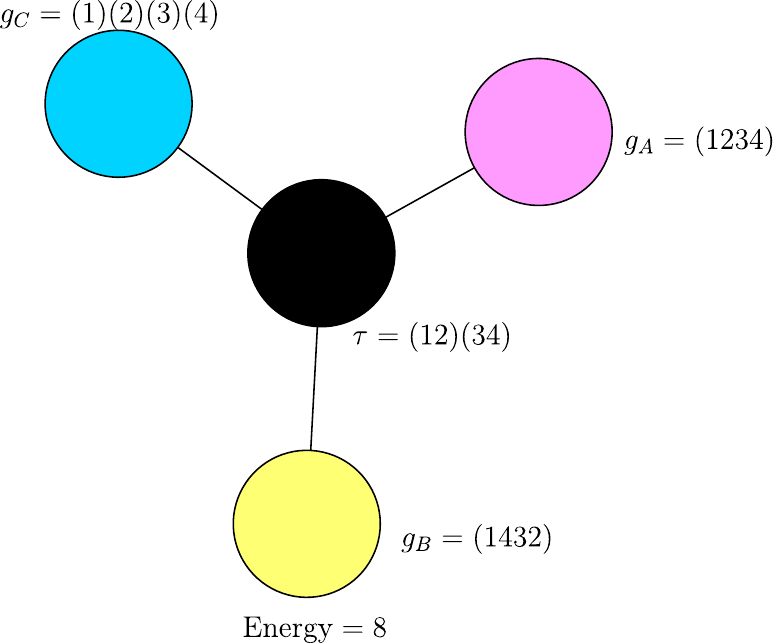}
 \caption{The minimum bulk configurations on the star graph for the $3$ and $4$-replica negativity. The central bulk vertex carries an intermediate permutation $\tau$ thereby breaking replica symmetry.}
 \label{fig:star-negativity}
\end{figure}

However, when we pick $(g_A, g_B, g_C)$ to be the ones that define the negativity: $g_A = (123)$, $g_B = (132)$, and $g_C = (1)(2)(3)$, then we find that replica symmetry is broken. Similarly, picking $g_A = (1234)$, $g_B = (1432)$ and $g_C = (1)(2)(3)(4)$, it again breaks replica symmetry. See Fig. \ref{fig:star-negativity} for an illustration.

The star graph from Fig. \ref{fig:star} is the simplest nontrivial example. Let us turn to analyzing more nontrivial graphs. However, then the numerical analysis is rather limited because of the huge number of permutations to sample over. For example, the $n = 3$ R\'enyi entropy requires $9$ replicas leading to $9!$ possible permutations at each vertex. Any graph containing more than one bulk vertex, say $N_b > 1$ bulk vertex, is therefore impossible in practice to brute force over since one needs to evaluate $(9!)^{N_b}$ possibilities.

Thus, for the more nontrivial graphs we discuss below, we need to make an ansatz for the class of permutations to study. We choose to scan over the restricted class of permutations given by 
\begin{equation}\label{eq:restricted-perms}
    \tau = g_A^{r_a} g_B^{r_b} g_C^{r_c}, 
\end{equation}
where $r_a$, $r_b$, and $r_c$ are integers taking values in $\{0, 1, 2\}$. This offers a significant speedup as we replace $9!$ permutations at every vertex with just $27$ permutations. However, even this dramatic simplification can only be pushed to around $6$ bulk vertices using our current machine power. Therefore, our analysis can not completely rule out the possibility of replica symmetry breaking in these nontrivial graphs. The $n = 2$ R\'enyi multi-entropy and the entanglement negativity for a small number of replicas are still tractable. We brute force over them whenever possible.

\subsubsection{Trivalent hyperbolic tree}

\begin{figure}
    \centering
    \includegraphics[width=0.5\linewidth]{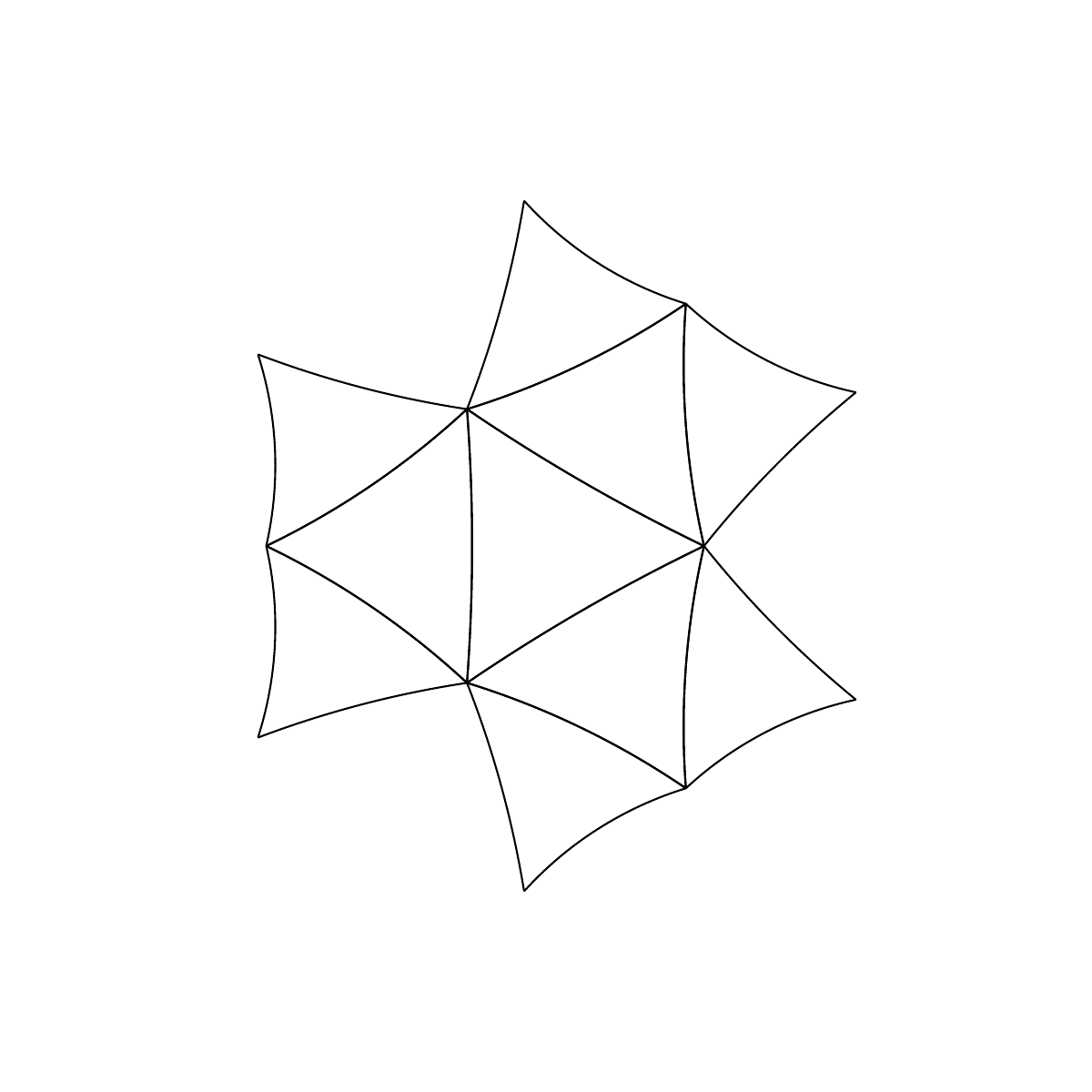}
    \hfill 
    \includegraphics[width=0.4\linewidth]{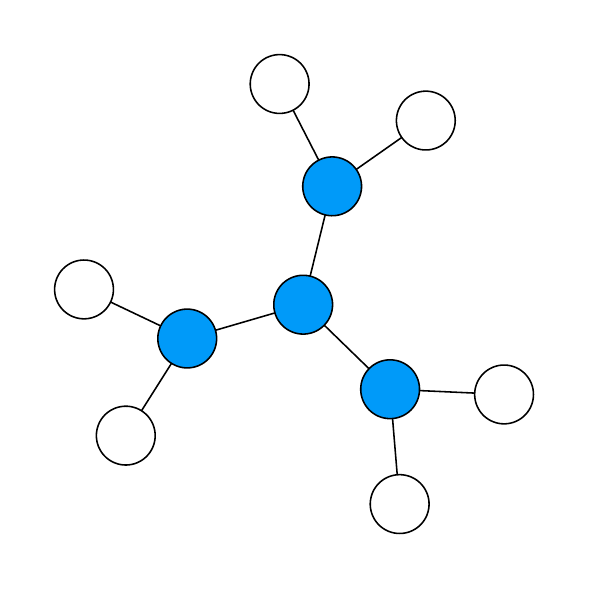}
    \caption{The $(3, 7)$ tessellation of the hyperbolic disk truncated at depth $2$ and its dual graph on which we setup our gauged random tensor network.}
    \label{fig:tri-tree-2}
\end{figure}

The next nontrivial example we consider is a trivalent hyperbolic tree shown in the right panel of Fig. \ref{fig:tri-tree-2}. This is the dual graph of the $(3, 7)$ tessellation of the hyperbolic disk truncated at depth $2$ shown in the left panel of Fig. \ref{fig:tri-tree-2}. Instead of one, we now have three bulk vertices to minimize over. As mentioned before, we cannot brute force over $9!^3$ due to computational limitations, and we stick to the restricted set of permutations for the $n = 3$ R\'enyi multi-entropy. For the $n = 2$ R\'enyi multi-entropy and the entanglement negativity, however, we consider the full set of permutations. The results of our analysis are as follows. 

For the $n = 2$ R\'enyi multi-entropy, we again see that the minimal energy configuration is replica symmetry preserving. For the $n = 3$ R\'enyi multi-entropy, the minimal energy configuration using the restricted class of permutations is also replica symmetry preserving. The permutations that define negativity, however, continue to break replica symmetry; see Figs. \ref{fig:(3,7,3)-best-multi} and \ref{fig:(3,7,3)-best-negativity}. 

\begin{figure}
    \centering
    \includegraphics[width=0.49\linewidth]{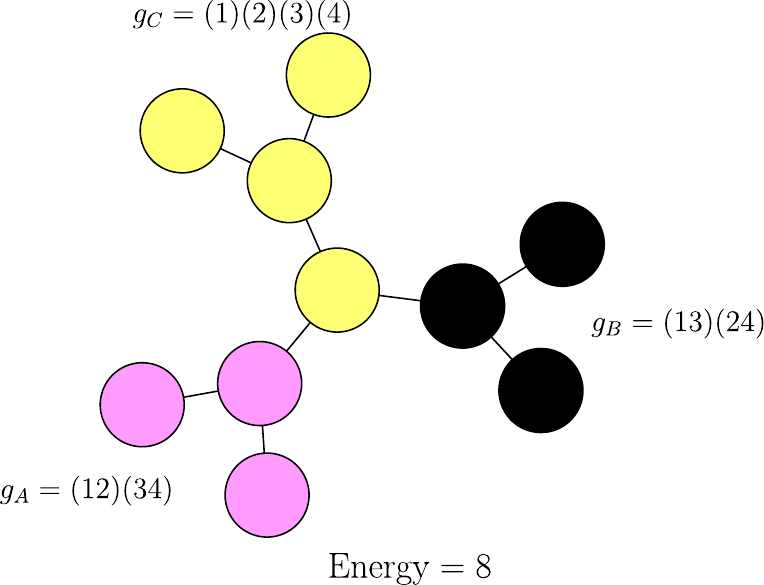}
    \hfill 
    \includegraphics[width=0.49\linewidth]{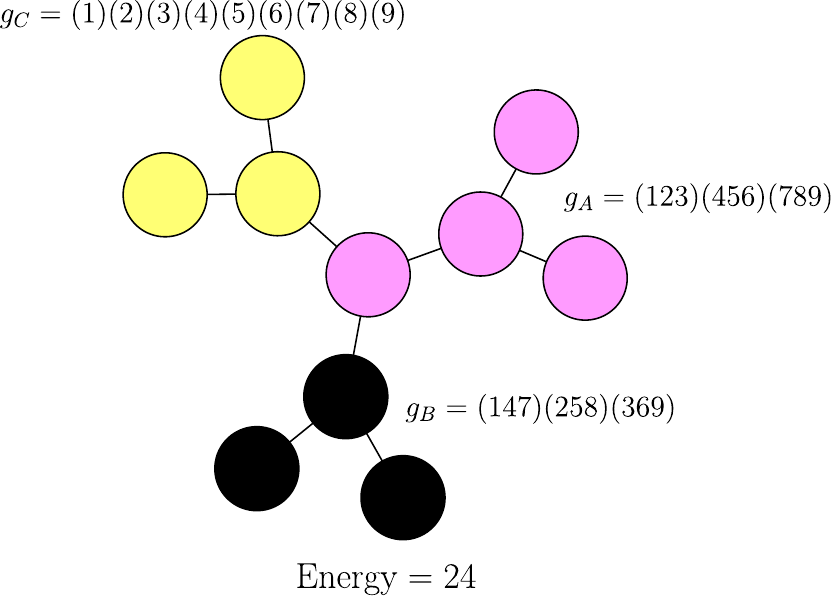}
    \caption{The minimal bulk configurations on the trivalent hyperbolic tree for the $n = 2$ and the $n = 3$ R\'enyi multi-entropy. The central bulk vertex in both cases carries the permutation $\tau = g_A$ thereby preserving replica symmetry.}
    \label{fig:(3,7,3)-best-multi}
\end{figure}

\begin{figure}
    \centering
    \includegraphics[width=0.49\linewidth]{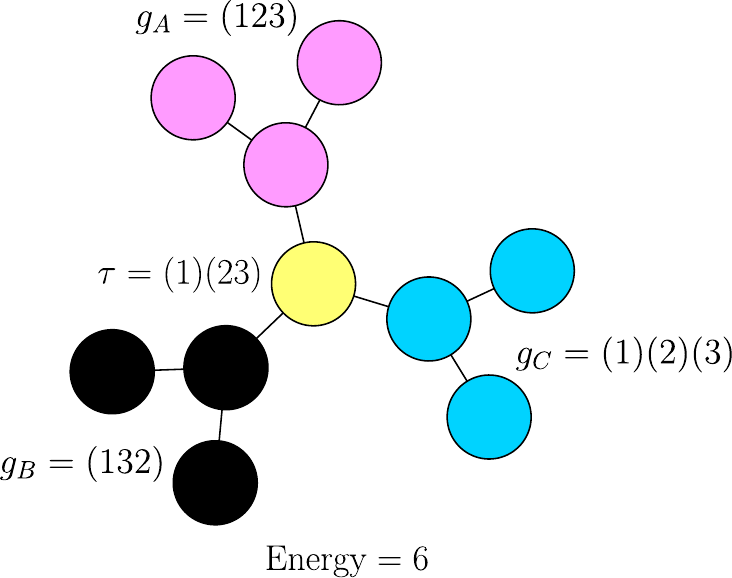}
    \hfill 
    \includegraphics[width=0.49\linewidth]{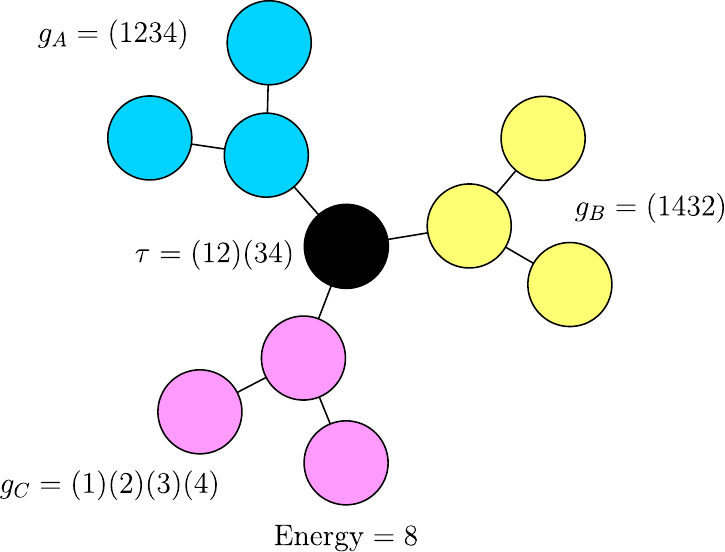}
    \caption{The minimal bulk configurations on the trivalent hyperbolic tree for the $3$ and $4$-replica negativity. The central bulk vertex in both cases is an intermediate permutation $\tau$ that breaks replica symmetry.}
    \label{fig:(3,7,3)-best-negativity}
\end{figure}

\subsubsection{The $(5,4)$ tessellation at depth 2}
The next example we consider is the $(5, 4)$ tessellation of the hyperbolic disk truncated at depth $2$ as shown in Fig. \ref{fig:(5,4,3)}. There are 11 bulk vertices and it is impossible to sample over all possible bulk configurations even with only $3$ replicas. To get around this problem, we partition the boundary edges such that all boundary vertices a given bulk vertex is connected to carry the same permutation. Under this choice, we assume that the permutation at this bulk vertex is fixed by the boundary permutation it is connected to. This is a valid assumption in conventional random tensor networks where the bond dimension is large and the bulk state is a tensor product of Bell states at every edge. We assume it is a reasonable approximation to make even with the gauge constraints to make the numerics tractable.

\begin{figure}
    \centering
    \includegraphics[width=0.49\linewidth]{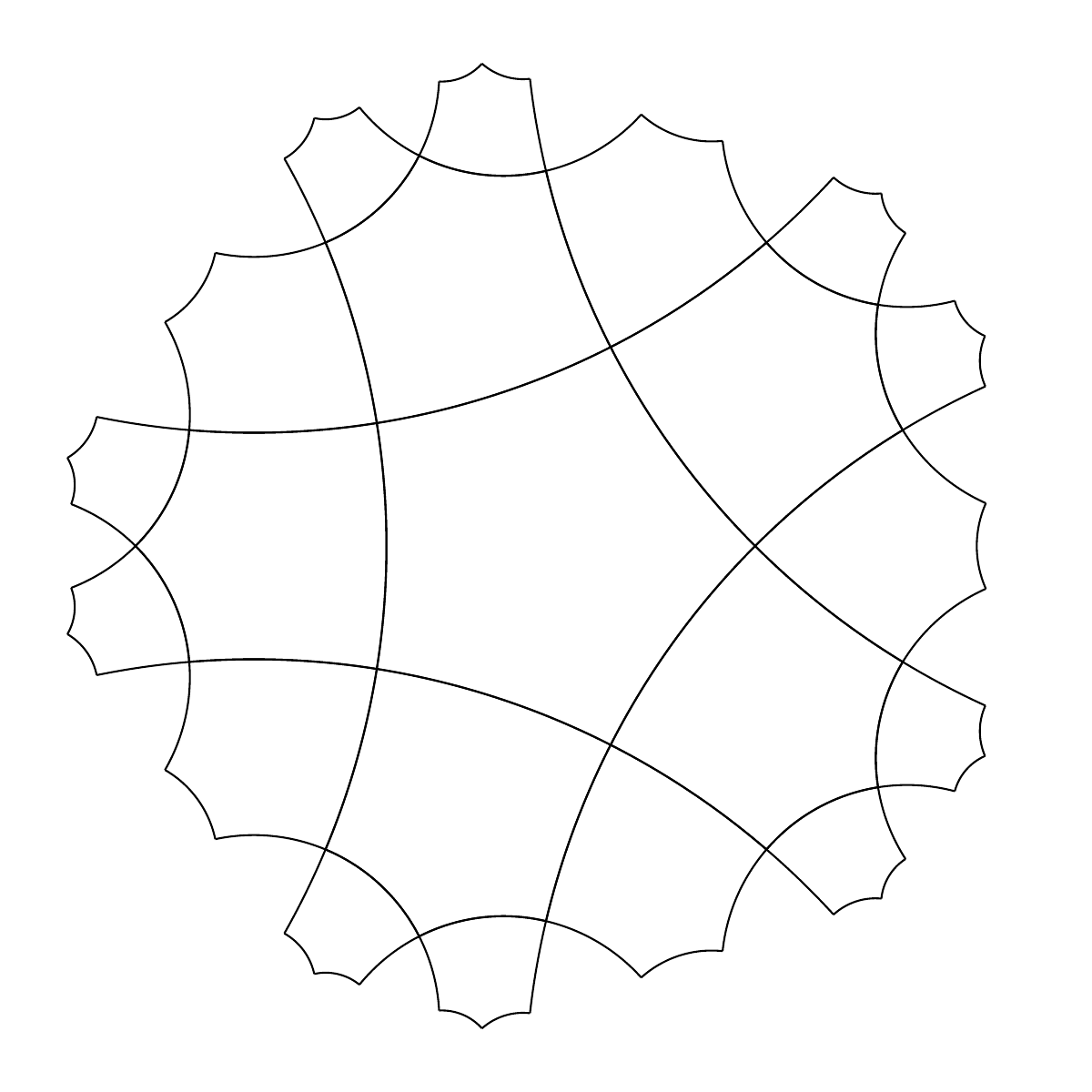}
    \hfill
    \includegraphics[width=0.49\linewidth]{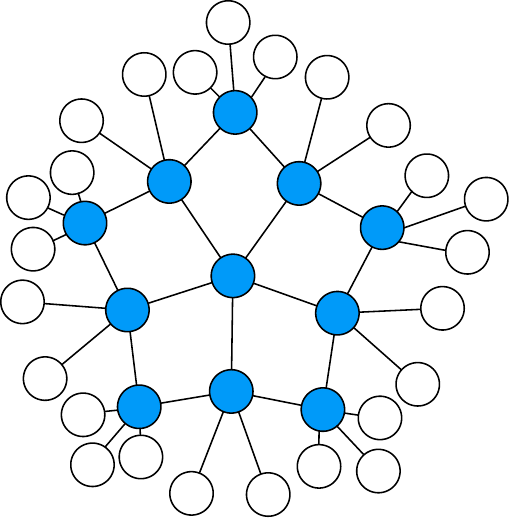} 
    \caption{The $(5,4)$ tessellation of the hyperbolic disk truncated at depth $2$ and its dual graph on which we setup the gauged random tensor network.}
    \label{fig:(5,4,3)}
\end{figure}

Under this assumption, we only need to minimize over the central bulk vertex which is quite tractable and can be brute forced over without restricting the set of permutations. The results are as follows. The 3-replica negativity, in this case, preserves replica symmetry while the $4$-replica negativity breaks it as shown in Fig. \ref{fig:(5,4,3)-best-negativity}. This could be because of the assumption we made. The global minima may be a different bulk configuration in this case. Coming to the R\'enyi multi-entropy, neither the $n = 2$ nor the $n = 3$ cases break replica symmetry as shown in Fig. \ref{fig:(5,4,3)-best-multi}. 

\begin{figure}
    \centering
    \includegraphics[width=0.45\linewidth]{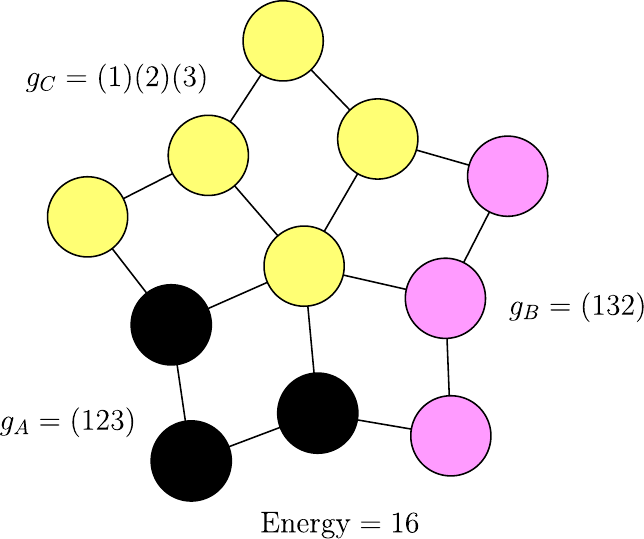}
    \hfill 
    \includegraphics[width=0.49\linewidth]{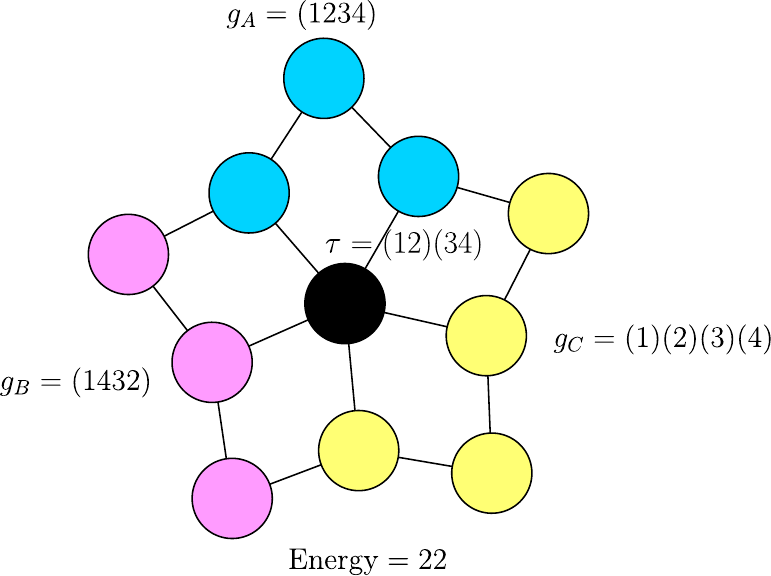} 
    \caption{The minimal bulk configurations for the $3$ and $4$-replica negativity on the $(5,4)$ tessellation of the hyperbolic disk. The $3$-replica case does not show replica symmetry breaking but the $4$-replica case does. This could be because we only scan over a restricted class of bulk configurations. }
    \label{fig:(5,4,3)-best-negativity}
\end{figure}

\begin{figure}
    \centering
    \includegraphics[width=0.45\linewidth]{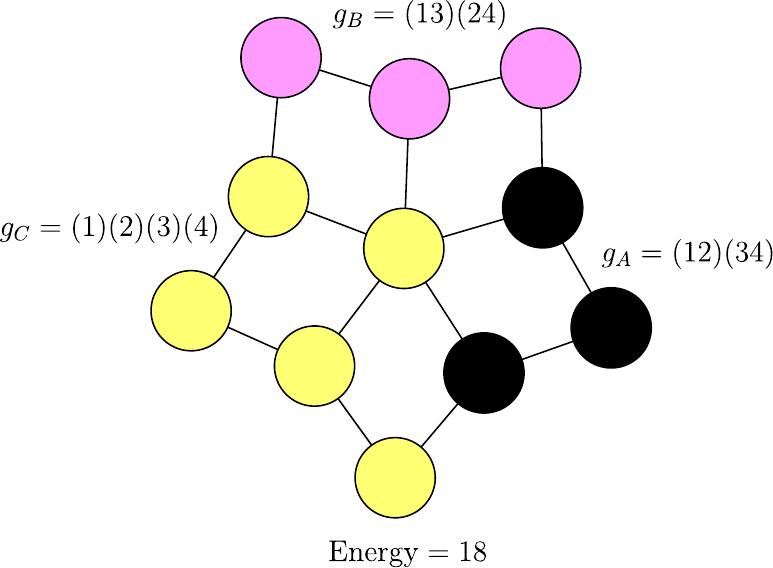}
    \hfill 
    \includegraphics[width = 0.54\textwidth]{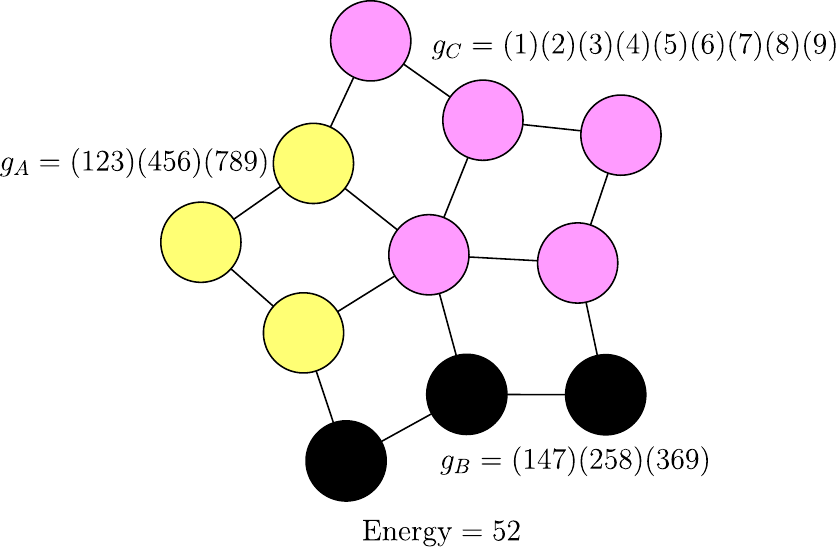}
    \caption{The minimal bulk configurations for the $n = 2$ and $n = 3$ R\'enyi multi-entropy on the $(5,4)$ tessellation of the hyperbolic disk. The central vertex carries the permutation $\tau = g_A$, thereby preserving replica symmetry.}
    \label{fig:(5,4,3)-best-multi}
\end{figure}

\subsection{Short summary}

In this section, we introduced a simple gauge extension of the conventional RTN domain-wall spin model and used it as a first probe of whether gauge structure can modify the replica pattern of multi-entropy.
More concretely, our proposal replaces the Bell-pair state on each internal edge by a gauge-invariant state from lattice gauge theory. When the gauge constraints are imposed, the resulting bulk state is no longer local across individual edges. Within the present toy $\mathbb{Z}_2$ model, our numerical results continue to show no evidence of replica symmetry breaking for the $n=2$ and $n=3$ R\'enyi multi-entropies, while negativity continues to exhibit RSB in the same setup.
This suggests that the no-RSB property of multi-entropy is more robust than one might have expected from the simplest ungauged RTN domain-wall spin model alone.
Broader discussion of the limitations of this toy construction and possible future directions will be given in Sec.~\ref{sec:discussion}.


\section{Summary and Discussion}
\label{sec:discussion}

It is useful to place our results in the broader context of the relation between RTN domain-wall spin models and holography.
The result established in Sec.~\ref{sec:noRSB} is a statement about the RTN domain-wall spin model itself, namely about the effective permutation-domain description associated with random tensor networks.
As such, it should not be interpreted as a direct no-go theorem for general holographic gravity duals.
Indeed, in the negativity case, the RTN/fixed-area analysis and the gravitational analysis point in the same qualitative direction: replica-symmetry-breaking saddles were first identified in the RTN setting~\cite{Dong:2021clv}, and were later shown to persist in general holographic states beyond the fixed-area limit~\cite{Dong:2024gud}.
The situation for multi-entropy is more subtle.
Our result shows that, within the conventional RTN domain-wall spin model, multi-entropy does not admit a nontrivial $\tau$-mediated replica-symmetry-breaking saddle.
At the same time, recent gravitational analyses suggest that the bulk replica structure of multipartite observables can be richer than what is captured by this spin-model description.
This is precisely where the conceptual tension lies.

From this viewpoint, our ungauged no-RSB result and the persistence of no RSB in the toy gauged model should be viewed together.
The former isolates a structural obstruction within the conventional RTN spin model: the boundary permutations relevant to multi-entropy do not admit a nontrivial common geodesic intermediate permutation $\tau$.
The latter shows that, at least in the present tripartite $n=2,3$ setup, this no-RSB property survives the introduction of a minimal $\mathbb{Z}_2$ gauge structure.
By contrast, negativity continues to exhibit RSB in the same gauged setup.
Taken together, these results suggest that multi-entropy is significantly less prone to replica symmetry breaking than negativity, and that this contrast is more robust than one might have expected from the simplest ungauged model alone.
At the same time, the scope of this gauged conclusion should be kept in mind.
The present construction is only a toy $\mathbb{Z}_2$ gauge extension, and therefore does not capture the large-bond dimension limit that underlies the usual RTN/domain-wall description.
Moreover, for some of the more nontrivial graphs, especially for the $n=3$ R\'enyi multi-entropy, our numerics relied on a restricted class of bulk permutations introduced in Eq.~(\ref{eq:restricted-perms}), rather than an exhaustive scan over all bulk configurations.
Our results should therefore be interpreted as evidence for the persistence of no RSB in this toy gauged setting, rather than as a complete classification of all possible gauged RTN replica saddles.

This perspective is also useful in comparing our results with previous RTN analyses of reflected entropy~\cite{Akers:2021pvd, Akers:2022zxr, Akers:2024pgq}. 
Those works show that random tensor networks can successfully capture certain multipartite geometric structures.
Our result suggests, however, that the conventional RTN domain-wall spin model is not universally sufficient for all multipartite observables.
In particular, even when multi-entropy admits a natural geometric interpretation from the holographic viewpoint, the corresponding RTN spin-model description does not produce a nontrivial $\tau$-mediated replica structure.
This indicates that the existence of a geometric multipartite picture does not guarantee that the conventional RTN spin model captures the full replica physics of the observable.

It is also important to clarify the relation to recent gravitational analyses of bulk replica symmetry, including the higher-replica discussion of Penington {\it et al.}~\cite{Penington:2022dhr} and the bulk-replica-symmetric multi-invariants studied in~\cite{Gadde:2024taa}.
These works address replica symmetry from the viewpoint of holographic handlebodies, orbifolds, and their possible symmetry properties in the bulk.
By contrast, the no-RSB statement established here concerns the absence of a $\tau$-mediated saddle in the RTN domain-wall spin model.
The two questions are therefore conceptually distinct.
Negativity provides an example where the RTN and gravitational perspectives point in the same qualitative direction, whereas multi-entropy exposes a more subtle situation.
In particular, our result suggests that the conventional RTN spin model does not fully capture the replica structures that may arise in gravitational treatments of multipartite observables, especially for multi-entropy.

At present, however, the precise origin of this discrepancy remains unclear.
It may be related to additional gravitational ingredients such as backreaction, topology change, gauge redundancy, or global constraints, but our results do not isolate which of these is decisive.
This uncertainty is tied in part to the limitations of the present toy construction itself.
When the gauge constraints are imposed, the resulting bulk state is non-local across edges, so the contribution of a given bulk configuration no longer factorizes edge by edge as in conventional random tensor networks.
Even at the numerical level, this significantly complicates the search over candidate replica saddles.
What they do show is that, for multi-entropy, the conventional RTN domain-wall spin model, even supplemented by the present toy gauge extension, does not reproduce all aspects of the replica structure that may be suggested by gravitational analyses.

The main lesson of this paper can therefore be summarized as follows.
Within the RTN domain-wall spin model, multipartiteness by itself does not imply replica symmetry breaking.
What matters is not merely that the observable is multipartite, but whether its boundary data admit a nontrivial common intermediate permutation that can mediate a competing saddle.
Negativity does, while multi-entropy does not.
Thus, in RTN language, the boundary data of multi-entropy are multipartite but not ``RSB-friendly'' in the same sense as those of negativity.

There are several directions for future work.
One obvious question is whether the robustness of no RSB for multi-entropy persists for more general gauge structures or more realistic holographically motivated extensions of RTNs.
In particular, it would be important to move beyond the present toy $\mathbb{Z}_2$ model and understand whether similar conclusions continue to hold in settings closer to the large-bond dimension regime of conventional random tensor networks.
Another is to understand more sharply how the conventional spin-model description should be enlarged if one wishes to capture gravitational replica phases associated with topology change or handlebody competition.
It would also be valuable to develop analytic control or more complete numerical methods that avoid the restricted scan over bulk permutations used here in some of the higher-replica examples.
It would also be interesting to clarify more systematically which multipartite observables are ``RSB-friendly'' and which are not, and whether this distinction can be characterized directly in terms of the structure of their boundary permutations.
More generally, one would like to understand whether the contrast between negativity and multi-entropy found here is specific to the present framework, or instead reflects a more general distinction among multipartite observables in holography. We hope to revisit these issues in future work.

\acknowledgments
We thank Shraiyance Jain for collaboration at an early stage of this project. We also thank Abhijit Gadde and Pratik Rath for their insightful suggestions. S.A. thanks the organizers of the Taiwan String Workshop 2025, where this collaboration was initiated. S.A. is supported by the Department of Atomic Energy, Government of India, under Project Identification No. RTI 4002, and the Infosys Endowment for the study of the Quantum Structure of Spacetime.
The work of N.I. was supported in part by MEXT KAKENHI Grant-in-Aid for Transformative Research Areas A “Extreme Universe” No. 21H05184. The work of N.I. was also supported in part by NSTC of Taiwan Grant Number 114-2112-M-007-025-MY3.

\appendix

\section{From random tensor networks to the domain-wall spin model}
\label{sec:app-crtn}

In this appendix, we briefly review the construction of conventional random tensor networks and the associated domain-wall description.

The setup is as follows.
We consider a graph $\Gamma=(V,E)$ consisting of a set of vertices $V$ connected by
edges $E$.
There is a set of boundary vertices $V_{\rm bdy}\subset V$ which have only one
incident edge, called the boundary edges $E_{\rm bdy}$.
To each edge, we associate a Hilbert space $\mathcal H$ of dimension $D$, and we
take the large-$D$ limit.
At every bulk vertex $x$, we choose a Haar-random state
\[
  \ket{T_x}\in \bigotimes_{y:(x,y)\in E}\mathcal H,
\]
where $y$ runs over all neighbors of $x$ in $\Gamma$.
For example, if the bulk vertex $x$ has three neighbors, then $\ket{T_x}$ is a
random state in $\mathcal H^{\otimes 3}$.

Picking such a random state at every bulk vertex produces the state
\[
  \bigotimes_{x\in V_{\rm bulk}}\ket{T_x}.
\]
The tensor network is a prescription for mapping this bulk state to a boundary
state defined on the Hilbert space of the boundary edges $E_{\rm bdy}$.
To each internal edge $(x,y)\in E_{\rm bulk}$ of the graph $\Gamma$, we associate
the maximally entangled Bell state
\begin{equation}
  \ket{\phi_{xy}}
  =
  \frac{1}{\sqrt D}\sum_{i=1}^D \ket{i_x}\ket{i_y}
  \in \mathcal H\otimes\mathcal H .
\end{equation}
The random tensor network then prepares the state
\begin{equation}
  \ket{\Psi}
  =
  \left(
    \bigotimes_{(x,y)\in E_{\rm bulk}}\bra{\phi_{xy}}
  \right)
  \left(
    \bigotimes_{x\in V_{\rm bulk}}\ket{T_x}
  \right)
\end{equation}
on the dangling boundary edges.

\begin{figure}[t]
  \centering
  \includegraphics[width=0.5\textwidth]{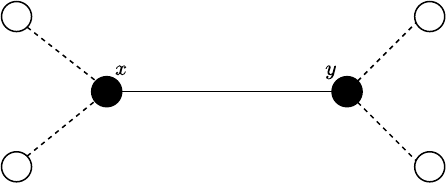}
  \caption{A simple random tensor network preparing a boundary state on four copies
  of the edge Hilbert space.}
  \label{fig:Gamma}
\end{figure}

As a simple example, consider the graph $\Gamma$ shown in Fig.~\ref{fig:Gamma}.
The boundary edges $E_{\rm bdy}$ are dashed and the boundary vertices
$V_{\rm bdy}$ are hollow.
For the two bulk vertices $V_{\rm bulk}=\{x,y\}$, we pick two random states
$\ket{T_x}$ and $\ket{T_y}$ from $\mathcal H^{\otimes 3}$.
For the bulk internal edge connecting $x$ and $y$, we pick the Bell state
$\ket{\phi_{xy}}$, and the boundary state prepared by the network is
\begin{equation}
  \ket{\Psi}
  =
  \bra{\phi_{xy}}
  \left(\ket{T_x}\otimes\ket{T_y}\right),
\end{equation}
which belongs to four copies of $\mathcal H$.

We can now calculate the R\'enyi entropies of a boundary subregion.
Let $A$ be a subset of the boundary vertices and $\bar A$ the complementary set.
The reduced density matrix on the edges of $A$ is
\begin{equation}
  \rho_A = \Tr_{\bar A}\ket{\Psi}\bra{\Psi}.
\end{equation}
The R\'enyi entropy is computed from
\begin{equation}
  \Tr \rho_A^n
  =
  \Tr\!\left[
    \left(\rho\otimes\cdots\otimes\rho\right)\mathcal C_A^{(n)}
  \right],
\end{equation}
where $\rho=\ketbra{\Psi}$ and $\mathcal C_A^{(n)}$ is the permutation operator
implementing the cycle $(1\,2\,\dots\,n)\in S_n$ on $n$ replicas of the Hilbert
space $\mathcal H_A$.

A key fact that allows one to compute the Haar average follows from the theory of
the symmetric subspace~\cite{Harrow:2013nib}:
\begin{equation}
  \overline{\left(\ket{T_x}\bra{T_x}\right)^{\otimes n}}
  \sim
  \sum_{\pi\in S_n} P(\pi),
\end{equation}
where $P(\pi)$ is the operator implementing the permutation $\pi$ on $n$ copies
of the Hilbert space $\mathcal H_x$ in which $\ket{T_x}$ lives.
There is an overall normalization, which we suppress, ensuring that the trace is
one.
If we choose a basis $\{\ket I\}$ for the Hilbert space $\mathcal H_x$, then
$P(\pi)$ acts as
\begin{equation}
  P(\pi)\ket{I_1,I_2,\dots,I_n}
  =
  \ket{I_{\pi(1)},I_{\pi(2)},\dots,I_{\pi(n)}}.
\end{equation}

\begin{figure}[t]
  \centering
  \includegraphics[width=0.45\textwidth]{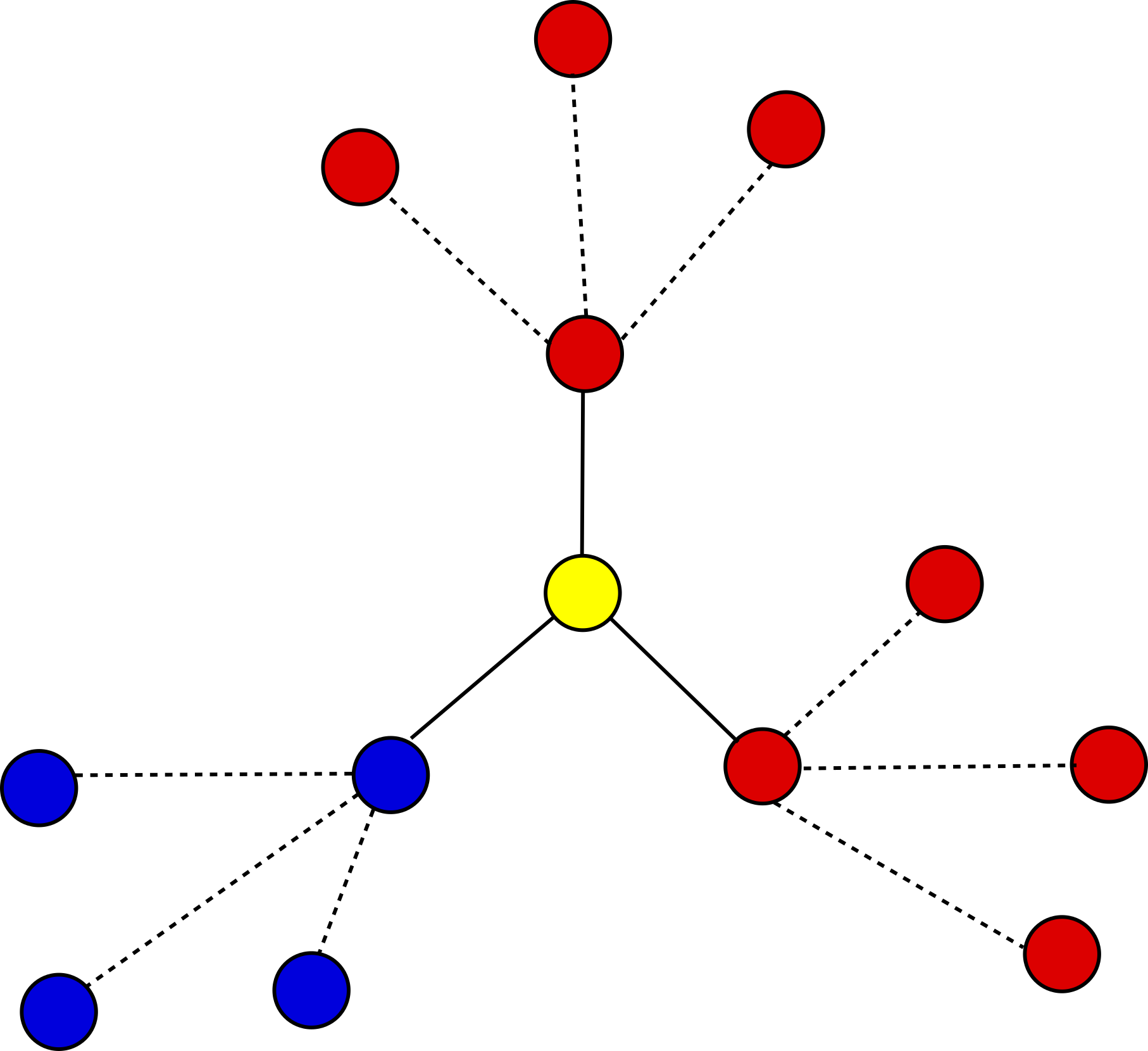}
  \caption{A particular bulk configuration where each color represents a different
  permutation of $S_n$. This is a tripartite multi-invariant evaluated on three
  Bell states living on the three internal edges.}
  \label{fig:ex-config}
\end{figure}

Using this fact about the symmetric subspace, we can rewrite
$\Tr\rho_A^n$ as
\begin{equation}
  \overline{\Tr\rho_A^n}
  \sim
  \sum_{\{\pi_x\}}
  \Tr\!\left[
    \left(
      \bigotimes_{(x,y)\in E_{\rm bulk}}
      \ket{\phi_{xy}}\bra{\phi_{xy}}^{\otimes n}
    \right)
    \left(
      \bigotimes_{x\in V_{\rm bulk}} P(\pi_x)
    \right)
    \mathcal C_A^{(n)}
  \right].
\end{equation}
The interpretation of each term in the above sum is as follows.
We first choose a configuration of permutations $\{\pi_x\}$ for the bulk vertices.
On the boundary vertices, the permutations are fixed to be the cycle
$\tau=(1\,2\,\dots\,n)$ in region $A$ and the identity
$(1)(2)\cdots(n)$ in the region $\bar A$. This choice of permutations defines a set of domain walls in the bulk of the
tensor network, and we are computing a multi-invariant~\cite{Gadde:2024taa}
of the Bell states on the internal edges.
See Fig.~\ref{fig:ex-config} for a simple example.
This observation already appeared in~\cite{Harlow:2016vwg}, although the language
of multi-invariants was introduced later.
Therefore, the sum above runs over all multi-invariants of a collection of Bell
states subject to a specific choice of permutations on the boundary vertices.
Different choices of boundary permutations and subregions compute different
multipartite entanglement measures of the boundary state.

In the large-$D$ limit, it is costly to introduce additional domain walls, and
the leading contribution to the above sum is given by the minimal cut subject to
the boundary conditions.
The dominant contribution is therefore
\begin{equation}
  \overline{\Tr\rho_A^n}\approx D^{(1-n)|\gamma|},
\end{equation}
where $|\gamma|$ is the length of the minimal cut.
Taking the logarithm and dividing by $1-n$ gives a flat R\'enyi entropy that
matches the Ryu--Takayanagi formula.
Since the state along the internal edges is maximally entangled, every edge gives
a contribution of $D^{1-n}$, resulting in a flat entanglement spectrum.
More generally, an edge connecting a vertex carrying the permutation $\sigma$ to
another carrying the permutation $\pi$ contributes a factor
$D^{-d(\sigma,\pi)}$, where $d(\sigma,\pi)$ is the Cayley distance.
It is defined as the minimum number of transpositions required to go from $\pi$
to $\sigma$.

Picking the Bell state for each internal edge accomplishes two things.
First, it gives a \emph{local} description of the bulk, since one can associate a
contribution of $D^{-d(\sigma,\pi)}$ to each edge that cuts across a domain wall.
Second, it allows one to recover the Ryu--Takayanagi formula. 

This is the conventional RTN/domain-wall construction that we modify in the main
text.
There, instead of feeding a collection of Bell states into the tensor network, we
feed a gauge-invariant state.
One appealing feature of that proposal is that, when no gauge constraints are
imposed, it reduces to the conventional RTN with Bell-pair contractions.
Once gauge constraints are imposed, however, the resulting bulk description is no
longer local in the same edge-by-edge sense.


\section{Lattice-gauge theory}
\label{sec:app-lgt}

In this appendix, we collect some details of the lattice-gauge-theory ingredients
used in Sec.~\ref{sec:gauge}.

\subsection{Basic lattice-gauge setup}

Let us begin with a brief review of lattice gauge theory.
The first ingredient of our setup is a finite group $G$ and the graph $\Gamma'$
obtained from the tensor network by discarding the boundary vertices and the
boundary edges.
To each edge, we assign an orientation and the Hilbert space $\mathcal{H}_e$
spanned by $\{\ket{g}: g \in G\}$ with inner product
\begin{equation}
  \langle g \hspace{-1mm}\mid  \hspace{-1mm} h \rangle  = \delta_{g,h}.
\end{equation}
The \emph{pre-gauged} Hilbert space is therefore
\begin{equation}
  \mathcal{H}_{\rm pre}=\bigotimes_{e\in E_{\rm bulk}} \mathcal{H}_e.
\end{equation}

In lattice gauge theory, there are two types of constraints to be imposed:
electric and magnetic~\cite{Kitaev:1997wr}.
We believe both are important for our purposes, but we only impose the electric
constraints for simplicity.
When we deal with the $\mathbb{Z}_2$ theory, the magnetic constraints are
somewhat implicit in our construction.
For more general gauge groups, this is no longer the case.

\begin{figure}[t]
    \centering
    \includegraphics[width=\textwidth]{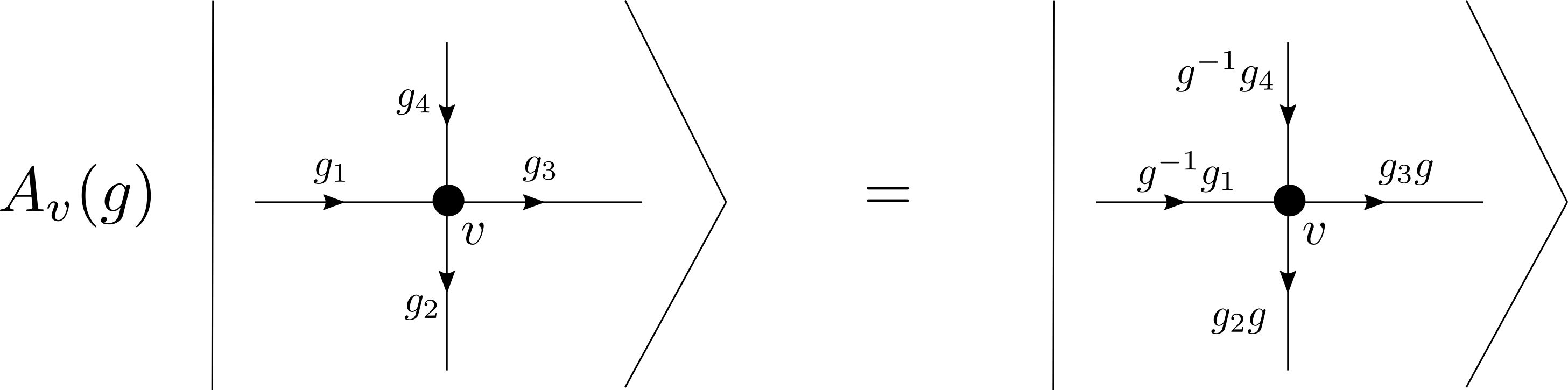}
    \caption{Action of the gauge transformation operator $A_v(g)$ on the edges
    incident on a bulk vertex $v$. Incoming edges are left-multiplied by $g^{-1}$
    and outgoing edges are right-multiplied by $g$.}
    \label{fig:Avg}
\end{figure}

The electric constraints are defined as follows.
We first define the operator $A_v(g)$, labelled by a bulk vertex $v$, acting on
all the edges incident on $v$.
The action of $A_v(g)$ is to left-multiply incoming edges by $g^{-1}$ and
right-multiply outgoing edges by $g$.
See Fig.~\ref{fig:Avg} for an example.

It is easy to verify that
\begin{equation}
  A_v(g)A_v(h)=A_v(gh).
\end{equation}
The electric constraint at the vertex $v$ is imposed by the operator
\begin{equation}
  A_v=\frac{1}{|G|}\sum_{g\in G} A_v(g),
\end{equation}
which is a projector, i.e.
\begin{equation}
  A_v^2=A_v.
\end{equation}
We impose this constraint at every vertex of the graph to define the physical
Hilbert space $\mathcal{H}_{\rm phys}$.
This is well defined because the operators $A_v$ and $A_w$ commute with each
other for distinct bulk vertices $v$ and $w$.
In equations,
\begin{equation}
  \mathcal{H}_{\rm phys}
  =
  \left(\bigotimes_{v\in V_{\rm bulk}} A_v\right)\mathcal{H}_{\rm pre}.
\end{equation}

\subsection{A simple \texorpdfstring{$\mathbb{Z}_2$}{Z2} example}

Let us consider a simple example of a $\mathbb{Z}_2$ gauge field with two vertices
connected by a single edge.
The pre-gauged Hilbert space is that of a qubit.
From Fig.~\ref{fig:Avg}, it is clear that $A_v(g)$ simply multiplies by the
Pauli $X$ operator if $g=1$ and does nothing if $g=0$.
Therefore, the electric constraint in this simple example is imposed by the
projector
\begin{equation}
  \frac{I+X}{2}.
\end{equation}
If we start with the state $\ket{0}$, which is the eigenstate of the Pauli $Z$
operator with eigenvalue $+1$, we end up with the gauge-invariant state
\begin{equation}
  \frac{\ket{0}+\ket{1}}{2}.
\end{equation}
More generally, for the case of a $\mathbb{Z}_2$ gauge group, the electric constraints are imposed by the star operators familiar from Kitaev's toric code \cite{Kitaev:1997wr}.

\subsection{Representation-theoretic interpretation}

There is a more physical interpretation of the electric constraint that goes as
follows.
From the representation-theoretic decomposition of $L^2(G)$, we have
\begin{equation}
  L^2(G)=\bigoplus_{\alpha}\mathcal{H}_\alpha\otimes \mathcal{H}^{\star}_{\alpha},
\end{equation}
where $\alpha$ labels irreducible representations of $G$.
The relation between the two bases is given by
\begin{equation}
  \ket{\alpha;ij}
  =
  \frac{\sqrt{d_\alpha}}{|G|}
  \sum_{g\in G} D^{\alpha}_{ij}(g)\ket{g},
\end{equation}
where $d_\alpha$ is the dimension of the irreducible representation $\alpha$
and $D^{\alpha}_{ij}$ are the matrix elements of the representation of $g$.
In this basis of irreducible representations of $G$, the electric constraint
imposes that the incoming and outgoing representations at the vertex fuse to
the identity.

\newpage

\begin{landscape}
\begin{center}
\vspace*{1.5cm}
\noindent\textbf{Cayley distance table for $S_4$}
\vspace*{0.2cm}
\label{app:S4distance}

\vspace{0.4em}

\scriptsize
\setlength{\tabcolsep}{2.0pt}
\renewcommand{\arraystretch}{1.0}
\resizebox{0.97\linewidth}{!}{%
\begin{tabular}{c|cccccccccccccccccccccccc}
  $d(g,h)$
  & $e$ & $(12)$ & $(13)$ & $(14)$ & $(23)$ & $(24)$ & $(34)$
  & $(12)(34)$ & $(13)(24)$ & $(14)(23)$
  & $(123)$ & $(124)$ & $(134)$ & $(234)$
  & $(132)$ & $(142)$ & $(143)$ & $(243)$
  & $(1234)$ & $(1243)$ & $(1324)$ & $(1342)$ & $(1423)$ & $(1432)$
  \\ \hline
  $e$
  & 0&1&1&1&1&1&1&2&2&2&2&2&2&2&2&2&2&2&3&3&3&3&3&3\\
  $(12)$
  & 1&0&2&2&2&2&2&1&3&3&1&1&3&3&1&1&3&3&2&2&2&2&2&2\\
  $(13)$
  & 1&2&0&2&2&2&2&3&1&3&1&3&1&3&1&3&1&3&2&2&2&2&2&2\\
  $(14)$
  & 1&2&2&0&2&2&2&3&3&1&3&1&1&3&3&1&1&3&2&2&2&2&2&2\\
  $(23)$
  & 1&2&2&2&0&2&2&3&3&1&1&3&3&1&1&3&3&1&2&2&2&2&2&2\\
  $(24)$
  & 1&2&2&2&2&0&2&3&1&3&3&1&3&1&3&1&3&1&2&2&2&2&2&2\\
  $(34)$
  & 1&2&2&2&2&2&0&1&3&3&3&3&1&1&3&3&1&1&2&2&2&2&2&2\\
  $(12)(34)$
  & 2&1&3&3&3&3&1&0&2&2&2&2&2&2&2&2&2&2&1&1&3&1&3&1\\
  $(13)(24)$
  & 2&3&1&3&3&1&3&2&0&2&2&2&2&2&2&2&2&2&3&1&1&1&1&3\\
  $(14)(23)$
  & 2&3&3&1&1&3&3&2&2&0&2&2&2&2&2&2&2&2&1&3&1&3&1&1\\
  $(123)$
  & 2&1&1&3&1&3&3&2&2&2&0&2&2&2&2&2&2&2&1&1&3&3&1&3\\
  $(124)$
  & 2&1&3&1&3&1&3&2&2&2&2&0&2&2&2&2&2&2&1&1&1&3&3&3\\
  $(134)$
  & 2&3&1&1&3&3&1&2&2&2&2&2&0&2&2&2&2&2&1&3&1&1&3&3\\
  $(234)$
  & 2&3&3&3&1&1&1&2&2&2&2&2&2&0&2&2&2&2&1&3&3&1&1&3\\
  $(132)$
  & 2&1&1&3&1&3&3&2&2&2&2&2&2&2&0&2&2&2&3&3&1&1&3&1\\
  $(142)$
  & 2&1&3&1&3&1&3&2&2&2&2&2&2&2&2&0&2&2&3&3&3&1&1&1\\
  $(143)$
  & 2&3&1&1&3&3&1&2&2&2&2&2&2&2&2&2&0&2&3&1&3&3&1&1\\
  $(243)$
  & 2&3&3&3&1&1&1&2&2&2&2&2&2&2&2&2&2&0&3&1&1&3&3&1\\
  $(1234)$
  & 3&2&2&2&2&2&2&1&3&1&1&1&1&1&3&3&3&3&0&2&2&2&2&2\\
  $(1243)$
  & 3&2&2&2&2&2&2&1&1&3&1&1&3&3&3&3&1&1&2&0&2&2&2&2\\
  $(1324)$
  & 3&2&2&2&2&2&2&3&1&1&3&1&1&3&1&3&3&1&2&2&0&2&2&2\\
  $(1342)$
  & 3&2&2&2&2&2&2&1&1&3&3&3&1&1&1&1&3&3&2&2&2&0&2&2\\
  $(1423)$
  & 3&2&2&2&2&2&2&3&1&1&1&3&3&1&3&1&1&3&2&2&2&2&0&2\\
  $(1432)$
  & 3&2&2&2&2&2&2&1&3&1&3&3&3&3&1&1&1&1&2&2&2&2&2&0
\end{tabular}%
}

\vspace{0.4em}
\captionof{table}{Cayley distance table for $S_4$ used in the main text.}
\label{tab:S4-distance}
\end{center}
\end{landscape}

\bibliography{refs}
\bibliographystyle{JHEP}

\end{document}